\documentclass[fleqn,usenatbib]{mnras}

\usepackage{newtxtext,newtxmath}
\usepackage[T1]{fontenc}

\DeclareRobustCommand{\VAN}[3]{#2}
\let\VANthebibliography\thebibliography
\def\thebibliography{\DeclareRobustCommand{\VAN}[3]{##3}\VANthebibliography}

\usepackage{graphicx}	% Including figure files
\usepackage{amsmath}	% Advanced maths commands
\usepackage{booktabs}
\usepackage{siunitx}   % For aligning numbers and uncertainties

\usepackage{multirow}  % For multirow cells
\usepackage{subfig}
\usepackage{makecell}
\usepackage{float} 
\usepackage{lineno}
\usepackage{placeins}
\usepackage{orcidlink}

\title[Bursty star formation in high$-z$ galaxies]{Signature of Bursty Star Formation in the High-Redshift Galaxies Detected with JWST}

\author[R. Sarkar \& S. Samui]{
Rupam Sarkar$^{1}$\orcidlink{0009-0000-7016-8085}\thanks{E-mail: rsarkar.astro@gmail.com}
and Saumyadip Samui$^{1}$\orcidlink{0000-0002-8052-0967}
\\
$^{1}$School of Astrophysics, Presidency University, 86/1 College Street, Kolkata, 700073, India
}

\date{Accepted XXX. Received YYY; in original form ZZZ}

\pubyear{\the\year{}}

\begin{document}
%\linenumbers   % Activate line numbers
\label{firstpage}
\pagerange{\pageref{firstpage}--\pageref{lastpage}}
\maketitle


\begin{abstract}
Recent JWST observations reveal an unexpectedly slow evolution in ultraviolet luminosity functions (UV LFs) at redshifts $z > 10$. To investigate this phenomenon, we develop a semi-analytical model of the UV LF, calibrated against well-constrained measurements at $z \sim 2-10$. Our analysis identifies a transition in star formation modes across cosmic epochs: at $z \lesssim 5$, a longer characteristic star formation timescale with nearly constant star formation efficiency ($f_\star$) dominates, whereas at $6 \lesssim z \lesssim 10$, shorter timescales prevail without requiring an increase in $f_\star$. For $z > 10$, the slow UV LF evolution is best explained by a shift toward even shorter star formation timescales without changing the star formation efficiency. Dust-free conditions or a top-heavy initial mass function (IMF) alone cannot reproduce the observations at $z\sim 14$. By combining UV LF with stellar mass estimates from \texttt{Prospector}-based SED fitting, we try to break degeneracies between IMF variations and star formation histories. Our results indicate that evolving star formation timescales rather than IMF or dust changes are the primary drivers of the observed high-redshift UV LF evolution, reflecting changing physical conditions during the earliest phases of galaxy assembly. Additionally, we show that moderate AGN activity could further boost UV luminosities at $z \sim 14$, potentially explaining the observed UV LF without changes in stellar parameters.
\end{abstract}


%%%%%%%%%%%%%%%%%%%%%%%%%%%%%%%%%%%%%%%%%%%%%%%%%%

%%%%%%%%%%%%%%%%% BODY OF PAPER %%%%%%%%%%%%%%%%%%

\section{Introduction}

%Over the years, 
Over the past few decades, the Hubble Space Telescope (HST) has enabled groundbreaking progress through its major blank-field observational surveys.
This includes the Hubble Ultra Deep Field (HUDF) survey \citep{Bouwens..2015,Livermore..2017}, Cosmic Assembly Near-infrared Deep Extra-galactic Legacy Survey (CANDELS) \citep{Finkelstein..2022a}, the Brightest of Reionizing Galaxyies (BoRG) survey \citep{BoRG..2011,BoRG..2015} and the Hubble Frontier Fields (HFF) survey \citep{Bouwens..2017,Bouwens..2022a,Atek..2018}.
Collectively, these efforts have identified thousands of galaxies at redshifts $z \geq 6 $
when the process of reionisation of the inter-galactic medium (IGM) is likely to be completed.
Even at redshifts as high as  $z\sim9-10$, a good number of galaxies was observed and we had a fare understanding of the galaxy population during this epoch \citep{Bouwens..2019,finklestein..2022,Bagley..2024}.
However, the sensitivity of HST is not sufficient to detect galaxies beyond $z\gtrsim10$, where the first galaxies, responsible for the beginning of reionisation, are formed.

Thanks to James Webb Space Telescope (JWST), now in the last few years, we have seen some extra ordinary and remarkable studies of these super-early galaxies which has revolutionized as well as challenged our understanding about the early universe.
Shortly after the commissioning of the JWST, several studies reported the existence of numerous bright ($m \leq 27.5$) galaxies at $z>10$ through photometric observations \citep[i.e.,][]{Castellano..2022,Naidu..2022,Perez..2023,Donnan..2023a,Donnan..2023b,Harikane..2023a,Robertson..2024,Adams..2024,Mcleod..2024,Finkelstein..2024}. Moreover, there are several spectroscopic confirmation of these high redshift (up to $z \sim 14.32$) candidates by the NIRSpec instrument \citep{Curtis..2023,Arrabal..2023,Bunker..2024,Carniani..2024}.

\par
These unexpected findings have posed significant challenges to earlier predictions from pre-JWST models and simulations, which apparently underestimated the abundance of such luminous galaxies at $z>10$ universe.
The high-redshift astrophysics community remains engaged in active debate regarding the potential explanations behind the overabundance of these bright galaxies, with proposed factors spanning a diverse array of hypotheses.
These include a top-heavy stellar initial mass function (IMF) \citep{Harikane..2023a,Finkelstein..2023, Finkelstein..2024}, variations in star-formation efficiency \citep{Dekel..2023,Mason..2023}, contribution from active galactic nuclei (AGN) \citep{Finkelstein..2024,Harikane..2023b,Fujimoto..2024}, reduced dust attenuation \citep{Ferrara..2023,Mason..2023}, and even the possible contamination of samples by lower-redshift interlopers \citep{Finkelstein..2024}.
Discussions also extend to fundamental cosmological considerations, such as potential deviations from standard cosmological models \citep{Liu..2022,Boylan..2023}. However, a consensus is yet to reach.

\par
In this study, we use a semi-analytical model to explore the potential drivers behind the observed slow evolution of the ultraviolet (UV) luminosity function (LF) for galaxies at redshifts $z > 10$.
We begin by examining the evolution of the UV LF across a broad redshift range ($z \sim 2-10$) to establish a baseline for comparison. 
We will show that the post-reionization universe ($z \lesssim 5$) tends to favor a longer characteristic star formation timescale, accompanied by an approximately constant star formation efficiency. 
In contrast, the early universe i.e. $6\lesssim z\lesssim 10 $, appears to support a shorter star formation timescale without a significant increase in star formation efficiency.
We then investigate such an evolution in the star formation time scale as seen at $z\le 10$ can solely be responsible for the slow evolution of the UV LF at $z > 10$ that has been observed.
Further, we investigate if observed/inferred stellar mass of these high redshift galaxies can be used to break the degeneracies between the already explored possibilities like top-heavy IMF or varying star formation efficiency with the evolving star formation time scale models.
To do that in a self consistent manner we also revisit the entire process of fitting different photometric fluxes of these high redshift galaxies (i.e. $z\ge 10$).
We carry out a detailed SED fitting using the Bayesian \texttt{Prospector} code and explore how photometric flux changes with different IMF, star formation time scale etc.

The paper is organized as follows.
In Section \ref{models}, we describe our semi-analytical models for the star formation mechanism and the UV luminosity function (LF).
In Section \ref{Fittings and observations}, we first constrain the UV LF model over the redshift range $z\sim 2-10$ using observed data points, and subsequently test its predictions at higher redshifts ($z\ge 10$).
In Section \ref{other models}, we explore alternative physical scenarios including dust attenuation, top-heavy initial mass functions, and potential AGN contribution to assess whether they can reproduce the observed UV luminosity function at $z\simeq 14$ and try to break the degeneracy among competing models.
In Section~\ref{prospector}, we employ the \texttt{Prospector} Bayesian framework \citep{Johnson..2021} to quantify how variations in the IMF and star-formation timescale affect the derived stellar mass and age of the $z_{\rm spec}=14.32$ galaxy, JADES-GS-z14-0 \citep{Carniani..2024}.
We assume a flat $\Lambda$-CDM cosmology throughout this paper with the cosmological parameters obtained from \cite{Planck..2018} observation, i.e., $\Omega_{\mathrm{m}} = 0.315, \Omega_{\Lambda} = 0.6847, \Omega_{\mathrm{b}} h^2 = 0.0224, \Omega_{\mathrm{CDM}} h^2 = 0.120$, $\sigma_8 = 0.811$ and $H_{\mathrm{0}} = 67.4 ~\mathrm{km\,s^{-1}\,Mpc^{-1}}$. Here, $h = H_{\mathrm{0}} /( 100~\mathrm{km\,s^{-1}\,Mpc^{-1}})$.

\section{Model and Methodology} \label{models}
\subsection{Halo Mass Function}
\label{sec:Halo mass function} 

To construct the UV luminosity function, it is essential to understand the abundance and evolution of dark matter haloes across cosmic time, since galaxies are assumed to form within these haloes.
The dark matter halo mass function (HMF), derived either analytically or through N-body simulations, provides good description of these haloes.
Note that, although simulations have made significant progress even at high redshifts $z \gtrsim 10$ \citep[i.e. MillenniumTNG,][]{MillenniumTNG}, they are still limited by the subgrid physics, and the enormous computational resource they need, restrict them to explore the vast parameter space related to star formation and associated feedback processes. On the other hand, our semi-analytical approach needs much less computational power and can be used to explore these parameter spaces easily.
In this work, we adopt the Sheth–Tormen (ST) formalism \citep{Sheth-Tormen..1999} to compute the HMF, as it shows good agreement with results from various simulations at redshifts $z < 10$. 
According to the ST formalism, the comoving number density of haloes with mass between $M$ and $M+dM$ at redshift $z$ is given by:

\begin{equation}
\begin{split}\label{e1}
    N_{\text{ST}}(M,z)\,dM &= \frac{\rho_0}{M}\left|\frac{1}{\sigma}\frac{d\sigma}{dM}\right| A\sqrt{\frac{2a}{\pi}} \\
    &\quad \times \left[1 + \left(\frac{\sigma^2}{a\delta_c^2}\right)^p\right] \frac{\delta_c}{\sigma} \exp\left(-\frac{a\delta_c^2}{2\sigma^2}\right)\,dM,
\end{split}
\end{equation}
where, $A = 0.3222$, $a = 0.707$, $p = 0.3$.
The quantity $\rho_0$ denotes the comoving average matter density of the universe, and $\delta_c=1.684$ is the linear critical density contrast for spherical collapse.
The variable $\sigma$ represents the root-mean-square (rms) of the linear density fluctuations, and is a function of both halo mass and the redshift at which collapse occurs. 
In our model, we compute the value of $\sigma(M,z)$ using the publicly available Code for Anisotropies in the Microwave Background \citep[\texttt{CAMB,}][]{Lewis..2011}.

\subsection{Star Formation Model} \label{sec-sfr}
After the collapse of the dark matter halo, the baryonic gas inside the halo is heated up to the virial temperature of the halo.
Due to radiative cooling, the gas cools down and starts accreting to the center of the halo.
Such accretion increases the baryonic density at the central part of the halo which leads to the star formation in a galaxy. 

We primarily adopt the star formation model proposed by \citet{Samui..2014}, in which the instantaneous star formation rate is assumed to be proportional to the available cold gas content within a galaxy.
This model additionally incorporates supernova feedback, radiative and AGN feedback and is constrained by observations of the UV luminosity functions of Lyman-break galaxies over the redshift range $1.5 \le z \le 10$. Further, this model can reproduce the observed correlations between star formation rate (SFR), stellar mass, and gas-phase metallicity, both in high redshift and nearby galaxies.
This model also reproduces the stellar to halo mass ratio in a wide halo mass range of $10^7 - 10^{13}$~M$_\odot$. The SNe feedback and AGN feedback that we assume here shape the stellar to halo mass ratio as revealed by observations, making our model more robust \citep[see][for more details]{Samui..2014}.
The baryonic mass outflow rate at time $t$ from the galaxy due to supernova (SNe) explosions is assumed to be proportional to the star formation rate at an earlier time $(t - t_{\text{SNe}})$, reflecting the delay between star formation and the onset of SNe.
Here, $t_{\text{SNe}}$ represents the characteristic timescale for supernova formation. Thus, the relationship is given by,
\begin{equation}
\dot{M}_w(t) = f_w\,\dot{M}_\star(t - t_{\text{SNe}}),
\end{equation}
where $f_w$ is the mass loading factor, $\dot{M}_w$ denotes the baryonic mass driven out of the host galaxy by SNe-induced wind and $\dot{M}_\star$ is the star formation rate.

The parameter $f_w$ depends on the outflow mechanism and is modeled as a function of the halo’s circular velocity $v_c$, following the relation $f_w = (v_c / v_c^*)^{-\alpha}$ \citep{Weaver..1977,Ostriker..1988,Veilleux..2005,Samui..2008}.
The exponent $\alpha$ takes the value of 2 for energy driven or cosmic ray driven outflows, and 1 for momentum-driven outflows, while $v_c^*$ is the characteristic circular velocity at which $f_w = 1$ \citep{Samui..2014}.

Setting $f_w = 0$ corresponds to a scenario with no supernova feedback. \citet{Samui..2014} demonstrated that the model with $\alpha = 2$ provides a good match to observational data for high-redshift galaxies. Therefore, in this work, we adopt $\alpha = 2$ along with $v_c^* = 100$ km/s.

When $t$ $\leq t_\text{SN}$ i.e. before the onset of SNe, the star formation rate is given by,
\begin{equation}
\frac{dM_\star(t)}{dt}  = \frac{f_\star M_b}{\kappa\tau} \frac{t}{\kappa\tau} \exp \left[ -\frac{t}{\kappa\tau} \right] \label{eq:t<t_sn}.
\end{equation}
Here, $M_b = ({\rm\Omega_b/\Omega_m)} M$ gives the total baryonic mass that has collapsed in a dark matter halo of mass $M$ and $f_\star$ denotes the star formation efficiency.
Further, $\tau$ is the dynamical time scale of the halo and $\kappa$ is a parameter that controls the duration of star formation activity in a given halo.
A higher value of $\kappa$ indicates a longer star formation time scale and a smaller value corresponds to a bursty mode of star formation.
For $t > t_\text{SN}$, one has to numerically solve the following delayed differential equation to get the star formation rate \citep[see][for details]{Samui..2018}:
\begin{equation}
\frac{d^2 M_\star(t)}{dt^2} = \frac{1}{\kappa\tau} \left[f_\star \frac{M_b}{\kappa\tau} e^{- \frac{t}{\kappa\tau}} - \frac{dM_\star(t)}{dt} - f_w \frac{dM_\star(t-t_\text{SN})}{dt}\right], \label{eq:t>t_sn}
\end{equation}
with initial conditions at $t=t_\text{SN}$
provided by the equation \ref{eq:t<t_sn}.

\par The star formation prescription given in equations \ref{eq:t<t_sn} and \ref{eq:t>t_sn} only accounts for the initial burst of star formation which is triggered by the halo assembly.
It does not take into account the passive star formation phase driven by slow gas accretion observed in the latter stages of the red sequence galaxies \citep{vdb..2008}. 
The initial burst
mode of star formation is assumed when the dark matter halo hosting the galaxy
collapsed from an over-dense region. This star formation approximately peaks in a dynamical time (depending on the value of $f_w$) of the halo and falls off exponentially
after that. With such a star formation scenario the galaxy will continue to shine
for a few dynamical time scale. By this time cold mode of gas accretion is likely
to take over to increase the available gas and continue a slow star formation in the galaxy.
To incorporate this latter process, we fix a constant star formation rate for galaxies of mass $M$ after their initial starburst, when it decreases to a value of $(M/10^{12})~ \rm M_\odot$/yr.
Thus, a $10^{10} ~\rm M_\odot$ halo would sustain a steady SFR of 0.01 $\rm M_\odot$/yr during its late evolutionary stages.
This normalization is motivated by the fact that our Milky Way with a halo mass of $10^{12} ~\rm M_\odot$ shows a sustained SFR about $1~\rm M_\odot$/yr \citep{Robitaille..2010}.
In our framework, the passive mode activates only after several dynamical timescales after the initial starburst.
\par The cooling efficiency of the gas in a given halo also plays a critical role in the star formation mechanism.
Due to the lack of heavier elements in the early universe, recombination line cooling from hydrogen and helium lines is favored most.
However, such cooling is only possible above a virial temperature ($T_{\text{vir}}$) of approximately \( 10^4 \)~K \citep{Barkana_Loeb..2001}. Therefore, we assume that gas in haloes with $ T_{\text{vir}} > 10^4 $~K can cool to form stars.
This condition establishes a lower mass threshold for star-forming haloes across cosmic time.

Further, photoionization of the intergalactic medium by UV photons increases the temperature of the gas which increases the Jean’s mass for collapse. This prevents low mass galaxies from collapsing and forming stars further \citep{Bromm-loeb..2002,Benson..2002,Dijkstra..2004,Samui..2008}. In order to model this photoionisation feedback,
we assume a complete suppression of star formation in haloes with $v_{c} \leq$ 35 km/s and no suppression for $v_{c} \geq$ 110 km/s.
For intermediate mass haloes we adopt a linear fit from 1 to 0 as the suppression factor.
Further, we consider a self-consistent reionization history of the universe that is constraint by the electron scattering optical depth ($\tau_e$) reported in \citet{Planck..2018}.

Moreover, in high-mass galaxies, AGN-driven feedback is expected to suppress the star formation mechanism.
To include this negative feedback in our model, we consider a suppression factor of $[1 + (M/10^{12} \rm M_\odot )^3]^{-1}$ on star formation in high mass haloes \citep{Bower..2006,Best..2006}.

\label{sec:Star formation model} % used for referring to this section from elsewhere

\subsection{UV Luminosity Function}
The star formation rate obtained from equations~\ref{eq:t<t_sn} and \ref{eq:t>t_sn} is converted to UV luminosity as follows.
For a given IMF of stars, we use the publicly available Flexible Stellar Population Synthesis (\texttt{FSPS}) package \citep{Conroy..2009,Conroy..2010} to get the UV luminosity, $l_{1500}(t)$ at 1500 Å as a function of time ($t$) if in total one solar mass of stars is formed in a single instantaneous burst at $t=0$ and continues to shine as per their life-time.
Within \texttt{FSPS} we use the Modules for Experiments in Stellar Astrophysics Isochrones
and Stellar Tracks (\texttt{MIST}) \citep{Dotter..2016} along with various choices of initial mass functions (IMFs) with a fixed metallicity of $\text{log}(Z/Z_\odot) \sim -1.5$ where $Z_\odot=0.0142$ \citep{Asplund..2009}.
We assume Salpeter initial mass function of stars in the mass range 1-100~$\rm M_{\odot}$ \citep{Salpeter..1955} in order to calculate the luminosity, $\ell_{1500}$.
The total UV luminosity of the galaxy is obtained from,
\begin{equation} \label{e3}
    L_{1500}(M,T) = \int_{T}^{0} \dot{M}_\star(M,T - t) \ell_{1500}(t) dt.
\end{equation}

But due to absorption by dust, only a fraction ($1/\eta$) of the total light produced by the stars can come out of the galaxy.
The UV luminosity ($L_0 = {L_{1500}}/{\eta}$) is converted to a standard AB magnitude, $\rm M_{AB}$, using the equation \citep{Oke..1983},
\begin{equation}\label{e4}
    {\rm  M_{AB}}= -2.5 \log_{10} (L_0) + 51.60,
\end{equation}
where $L_0$ is in units of erg s$^{-1}$ Hz$^{-1}$.
Finally, the luminosity function $\Phi({\rm M_{AB}},z)$ at a given $z$ is obtained from \citep[see][]{Samui..2007},
\begin{equation} \label{e5}
    \Phi({\rm {M_{AB}}},z)  d{\rm M_{AB}} = \int_{z}^{\infty} dz_c \frac{d N_{ST}(M,z_c)}{dz_c} \frac{dM}{dL_{1500}} \frac{dL_{1500}}{d\rm M_{AB}} d{\rm M_{AB}}.
\end{equation}

\par
Note that the dust attenuation factor ($\eta$) decreases with increasing redshift \citep[i.e.][]{Bouwens..2012,Burgarella..2013,Khaire..2015}.
We adopt the LMC2 extinction model with the median parameter set from \citet{Khaire..2015} as our fiducial dust attenuation model, which is calibrated against observations and captures the expected decline of dust attenuation at high redshift.
The attenuation factor is computed as $\eta = 10^{A_\mathrm{FUV}/2.5}$, where $A_\mathrm{FUV}$ is given by
\begin{equation}\label{e6}
    A_\text{FUV}(z) = \frac{1.42+0.93z}{1+(z/2.08)^{2.2}}.
\end{equation}
The values of $\eta$ obtained using the high and low parameter sets (i.e. $1\sigma$ uncertainty) of the LMC2 extinction model lie within 10\% of those derived from the median parameter set for $z < 10$. Furthermore, the median parameter values of other commonly used extinction models (i.e. SMC, LMC, and Calzetti) also lie within this same range over the corresponding redshift interval. Therefore, we adopt this extinction model as our fiducial choice for the analysis.

We fit the observed UV luminosity function at a given redshift with the $\chi^2$ minimization technique in order to constrain the star formation efficiency parameter, $f_\star$.
Note that, $f_\star$ does not precisely correspond to the traditional definition of star formation efficiency used in the literature.
The total baryonic mass ultimately converted into stars is given by $M_bf_\star/(1 + f_w)$ (see \cite{Samui..2014} for a detailed calculation).

\begin{figure*}
\centerline{
\includegraphics[width=\textwidth]{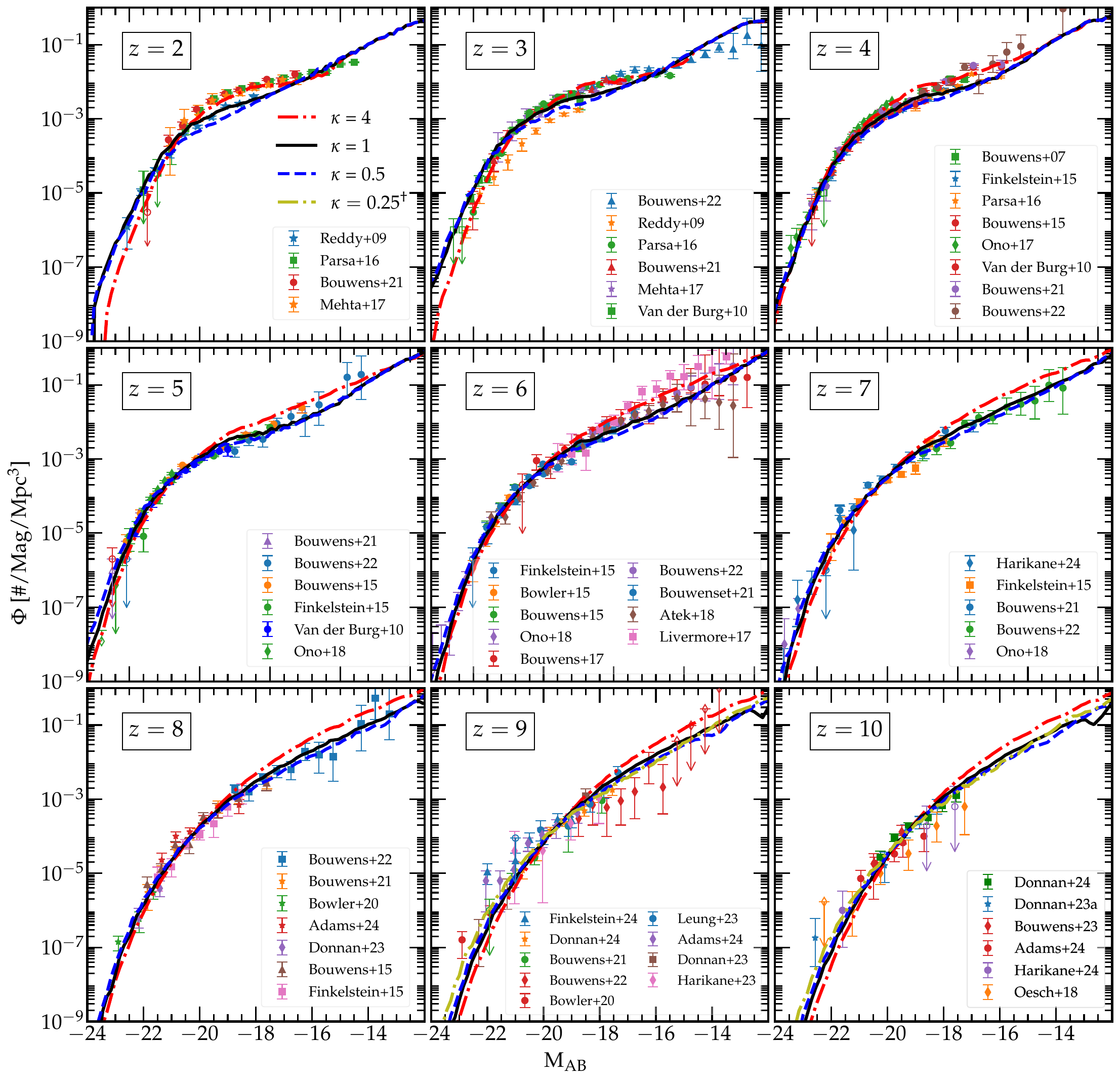}%

}
\caption{UV luminosity functions at $z=2-10$. Models are shown with $\kappa=0.5$ (thin dashed blue line), $\kappa=1$ (solid black line) and $\kappa=4$ (dash-dotted red line). Additionally model with $\kappa = 0.25$ ( thick dash-dotted olive line; marked with $\dagger$ in the $z=2$ panel ) is included for $z=9$ and $z=10$. Filled markers represent observational data points included in the fitting procedure, while unfilled markers indicate upper or lower limits (excluded from fitting). The observed data points are taken from \citet{Bouwens..2007,Reddy..2009,vdb..2010,Bouwens..2015,Bowler..2015,Finkelstein..2015,Parsa..2016,Bouwens..2017,Mehta..2017,Livermore..2017,Ono..2018,Oesch..2018,Atek..2018,Bowler..2020,Bouwens..2021,Bouwens..2022a,Harikane..2023a,Donnan..2023a,Donnan..2024,Adams..2024,Harikane..2025}.}
\label{fig:LF:2-7}
\end{figure*}

\begin{table*}
\centering
\begin{tabular}{cccccccccc}
\toprule
\multirow{2}{*}{$z$} & \multirow{2}{*}{$\eta$} & 
\multicolumn{2}{c}{$\kappa = 4.0$} & \multicolumn{2}{c}{$\kappa = 1.0$} & 
\multicolumn{2}{c}{$\kappa = 0.5$} & \multicolumn{2}{c}{$\kappa = 0.25$} \\
\cmidrule(lr){3-4} \cmidrule(lr){5-6} \cmidrule(lr){7-8} \cmidrule(lr){9-10}
& & $f_\star$ & $\chi^2_\nu$ & $f_\star$ & $\chi^2_\nu$ & $f_\star$ & $\chi^2_\nu$ & $f_\star$ & $\chi^2_\nu$ \\
\midrule
2    & 4.834 & $0.304\pm0.014$ & {20.11}$^\dagger$ & $0.179\pm0.014$ & 58.79 & $0.111\pm0.010$ & 87.46 &-- & -- \\
3    & 3.311 & $0.301\pm0.010$ & {13.79}$^\dagger$ &  $0.146\pm0.007$ & 24.35 &  $0.099\pm0.003$ & 35.70 & -- & -- \\
4    & 2.479 & $0.325\pm0.001$ & {29.06}$^\dagger$ &  $0.094\pm0.001$ & 36.80 &  $0.057\pm0.001$ & 80.36 &-- & -- \\
5    & 2.032 & $0.276\pm0.008$ & 24.73 &  $0.079\pm0.002$ & 7.69 &  $0.055\pm0.001$ & {5.71}$^\dagger$ &-- & -- \\
6    & 1.771 & $0.257\pm0.007$ & 7.59 &  $0.065\pm0.002$ & {3.89}$^\dagger$ &  $0.041\pm0.001$ & 4.69 &-- & -- \\
7    & 1.605 & $0.276\pm0.014$ & 5.58 &  $0.066\pm0.003$ & 2.21 &  $0.040\pm0.002$ & {2.06}$^\dagger$ &-- & -- \\
8    & 1.493 & $0.316\pm0.015$ & 2.91 &  $0.067\pm0.001$ & 1.21 &  $0.039\pm0.001$ & {1.05}$^\dagger$ &-- & -- \\
9    & 1.413 & $0.309\pm0.020$ & 4.01 &  $0.066\pm0.003$ & 2.25 &  $0.039\pm0.001$ & 1.50 & $0.027\pm0.001$ & {1.18}$^\dagger$ \\
10   & 1.353 & $0.335\pm0.023$ & 2.36 &  $0.069\pm0.004$ & 1.36 &  $0.039\pm0.003$ & 1.08 & $0.026\pm0.001$ & {0.97}$^\dagger$\\ \hline
11   & 1.307 & $0.638\pm0.044$ & 3.44 &  $0.120\pm0.006$ & 1.86 &  $0.068\pm0.004$ & 1.36 & $0.042\pm0.001$ & {1.27}$^\dagger$ \\
12.5 & 1.256 & $0.725\pm0.090$ & 3.77 &  $0.137\pm0.015$ & 2.38 &  $0.078\pm0.006$ & {1.91}$^\dagger$ & $0.043\pm0.005$ & 1.99 \\
14   & 1.218 & $>1^{\ddagger}$ & -- &  $0.231\pm0.033$ & 1.66 &  $0.120\pm0.015$ & {1.53}$^\dagger$ & $0.064\pm0.007$ & 1.66 \\
\bottomrule
\end{tabular}
\caption{Comparison of the best-fit $f_\star$ values and the corresponding reduced chi-squared ($\chi^2_\nu$) values across the redshift range $z = 2$--$14$ for different $\kappa$ values. The $\kappa = 0.25$ case is evaluated only for $z \geq 9$. The lowest $\chi^2_\nu$ among all $\kappa$ models at each redshift is indicated by $\dagger$.The symbol $\ddagger$ denotes an unphysical value of $f_\star$ for the $\kappa = 4$ model at $z = 14$. }
\label{table:all}

\end{table*}

\begin{figure*}
\centerline{
\includegraphics[width=\textwidth]{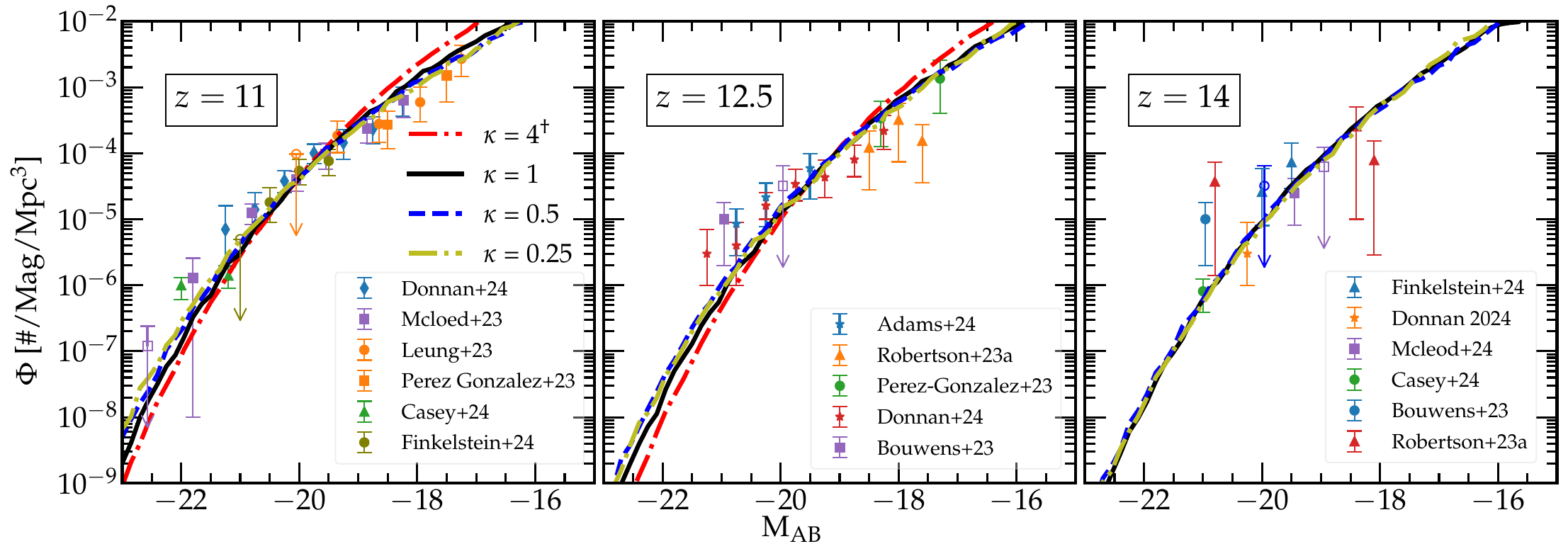}%

}
\caption[]{UV luminosity functions at $z=11-14$. Models are shown with $\kappa=0.25$ (thick dash-dotted olive line), $\kappa=0.5$ (thin dashed blue line), $\kappa=1$ (solid black line) and $\kappa=4$ (dash-dotted red line,$\dagger$ not included for $z=14$). Filled markers represent observational data points included in the fitting procedure, while unfilled markers indicate upper or lower limits (excluded from fitting). The observed data points are taken from \citep{Bouwens..2022a,Bouwens..2023,Leung..2023,Perez..2023,Donnan..2023a,Donnan..2024,Mcleod..2024,Casey..2024,Finkelstein..2024,Adams..2024,Robertson..2024,Harikane..2025}
}
\label{LF:11-14}
\end{figure*}

\section{Results} \label{Fittings and observations}

\subsection{UV Luminosity Functions at $z \le 10$}

We first show the robustness of our UV luminosity function (LF) model described in the previous section by comparing its predictions with observed data available up to redshift $z \sim 10$. Note that a substantial number of observed data points over the magnitude range $-24\lesssim \text{M}_{\rm AB}\lesssim -12$, are available which provide tight constraints on the faint end of the UV luminosity function in the redshifts range $2\le z \le 10$ \citep{Bouwens..2007,Reddy..2009,vdb..2010,Bouwens..2015,Bowler..2015,Finkelstein..2015,Parsa..2016,Bouwens..2017,Mehta..2017,Livermore..2017,Ono..2018,Atek..2018,Bowler..2020,Bouwens..2021,Bouwens..2022a,Harikane..2023a,Donnan..2023a,Donnan..2024,Adams..2024,Harikane..2025}.
The faint end of the UV LF is highly sensitive to baryonic feedback processes, including radiative and supernova-driven feedback, which suppress star formation in low-mass haloes and plays important role in constraining the feedback models.
As mentioned in the previous section, we vary the star formation efficiency, $f_\star$ to fit the model prediction with the observed UV LF using $\chi^2$ method \citep[similar to][]{Samui..2018}. 
For each redshift this fitting is repeated across a range of $\kappa$ values, which modulate the star formation timescale and significantly impact the shape the predicted UV LF.

Figure~\ref{fig:LF:2-7} shows the best-fit UV LFs at $ z = 2 - 10$ for models with $\kappa = 4$, 1 and 0.5 along with the observational data. For $z \geq 9$, we additionally include $\kappa = 0.25$ model to account for the burst-like star formation expected to dominate in the early universe, a scenario we explore further in the following section.
Table~\ref{table:all} lists the best-fit $f_\star$ values for each redshift and $\kappa$ models, along with the associated reduced chi-square statistics ($\chi^2_\nu$). Note 
that the reduced chi-square value may be higher due to the scatter in the observed data points themselves.
It is clear from the figure as well as from the $\chi^2$ statistics that our models provide a good description of the
observed UV luminosity functions in a wide range of magnitude, from the faint end $\rm M_{AB}\sim -12$ to the bright end
$\rm M_{AB}\sim -24$, which is five orders in terms of luminosity. Further, our feedback models along with the evolution
in the halo mass function reproduce the
evolution in the observed luminosity function from $z=2$ to $z=10$. A detailed investigation reveals even
more interesting trends. 
For instance, the model with $\kappa = 4$, corresponding to a relatively extended star formation timescale, maintains a nearly constant star formation efficiency ($f_\star\sim 0.3$) over the full redshift range $z \le 10$.
This implies a scenario of steady star formation activity in galaxies, consistent with the gradual accumulation of stellar mass over time.
In contrast, models with shorter star‐formation timescales ($\kappa = 1$ and $0.5$) in the post‐reionization epoch ($z \approx 2 - 5$), display a pronounced rise in $f_\star$, increasing by a factor of $\sim2.75$ between $z = 6$ and $z = 2$.
However, this increased efficiency comes at the cost of a poorer statistical fit to the observed LFs.
For example, at $z \approx 2$, the $\kappa = 4$ case yields $\chi^2_\nu = 20.11$, whereas $\kappa = 1$ and $\kappa = 0.5$ produce $\chi^2_\nu = 58.79$ and $87.46$, respectively, demonstrating that longer star formation timescales are statistically favored at $z \leq 5$.

Note that, at a fixed redshift and halo mass, galaxies with higher $\kappa$ require a larger $f_\star$ to achieve the same luminosity as those with lower $\kappa$. Consequently, models with lower $\kappa$ naturally need lower $f_\star$ values to reproduce the UV luminosity function. This behavior is discussed in more detail in Appendix~\ref{sfh_kappa_relation}. Further, changing the dust attenuation model to SMC, LMC, or Calzetti instead of LMC2 would only lead to a change of $\sim 10\%$ in the best fit values of $f_*$ without changing the minimum $\chi^2$ value. Also adopting a different metallicity value has a tiny effect on the resulting luminosity functions as discussed in Appendix~\ref{Effect of metalicity}.

Now, in the high-redshift regime (i.e. $z = 6 - 10$), the trend reverses.
Here, models with shorter star formation timescales, i.e., $\kappa = 1$ and 0.5 along with approximately constant $f_\star$ values of $0.067$ and $0.040$, respectively, yield significantly better fits to the observed UV LF data compared to the $\kappa = 4$ case.
For instance, at $z = 7$ the $\kappa = 4$ model produces $\chi^2_\nu = 5.58$, whereas the $\kappa = 1$ and $\kappa = 0.5$ cases give $\chi^2_\nu = 2.21$ and $2.06$, respectively.
These results strongly suggest a preference for shorter, more intense star formation episodes in early galaxies.
Thus we conclude that our feedback regulated star formation models provide a good fit to the observational data upto a redshift 10 and also reveal that the nature of star formation was more bursty in the past.
Therefore, we further investigate if such an evolution can be seen in newly observed galaxies at $z\ge 10$ by JWST.

\begin{figure}
\centering
\includegraphics[width=\columnwidth]{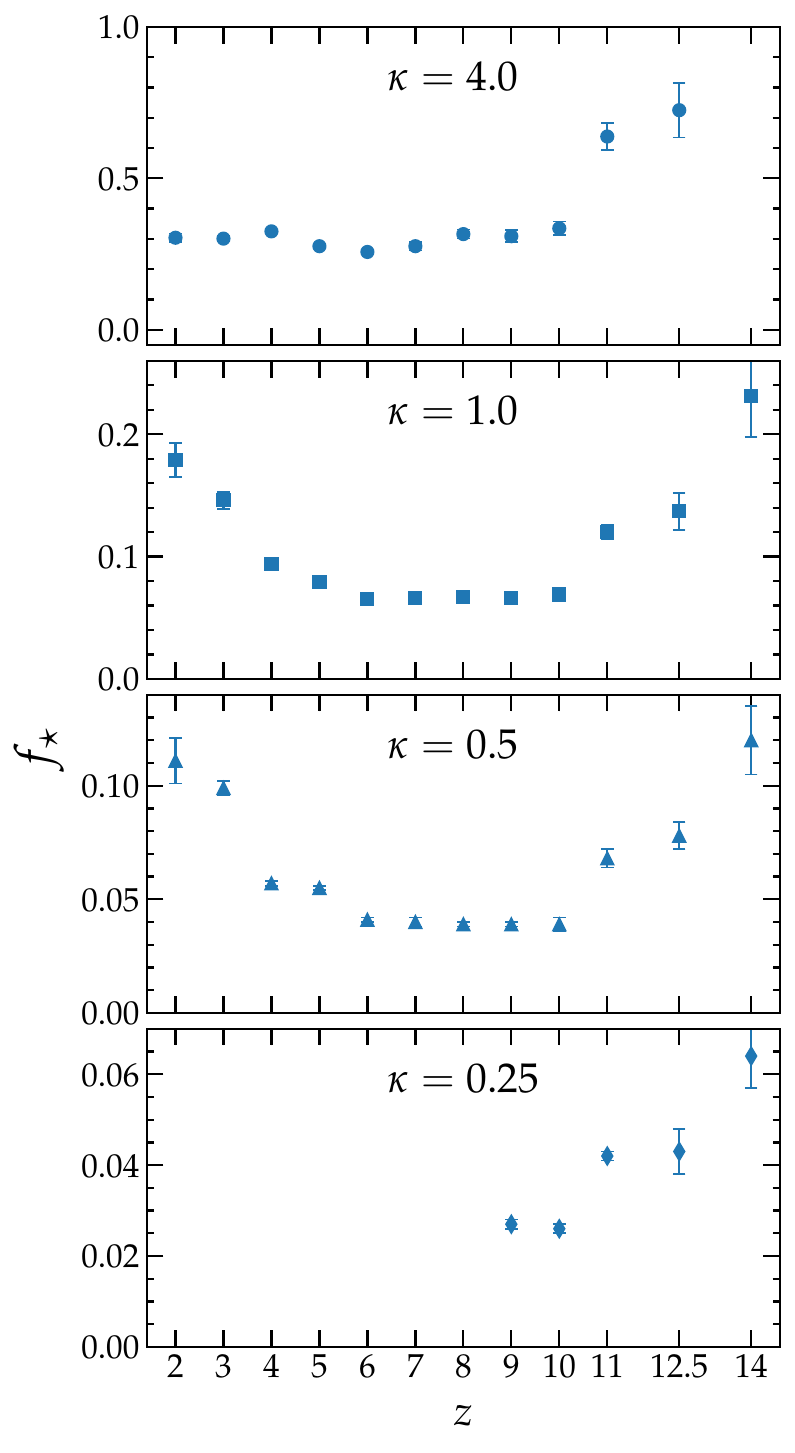}
\caption[]{The figure shows the redshift evolution of star formation efficiency ($f_{\star}$) along with its uncertainty for models with different $\kappa$, i.e., $\kappa=4,$ , 1 and 0.5. Additionally $\kappa=0.25$ is included for redshift 9-14 only.   }
\label{fig:f_star_vs_z}
\end{figure}

\subsection{UV Luminosity Function at $z>10$}

In the previous section we have seen that at redshifts $z\leq10$ observations reveal a smooth and continuous decline in the UV LF of galaxies with increasing redshift which can be understood as a moderate change in star formation time scale along with the evolution in the dark matter halo formation expected from standard $\Lambda$CDM cosmology (see Figure~\ref{fig:LF:2-7}). 
However at $z>10$, as revealed from recent observations by JWST, this evolutionary trend appears to decelerate, deviating from the expected decline extrapolated from lower redshifts. 
This observed departure suggests either a higher-than-anticipated number density of galaxies or an enhancement in their intrinsic UV luminosities during the earliest stages of cosmic history. 
In this section, we examine the underlying mechanisms that could be responsible for this apparent slowdown in the evolution of the UV LF. 

In Figure~\ref{LF:11-14} we have plotted the observed luminosity functions for $z> 10$ along with our model predictions. It is clear from the figure that our models
provide a good fit to the observed luminosity functions during the cosmic dawn.
Even though our models can explain the observed luminosity functions in these redshifts with a reasonably good chi-square (low values of chi-square are due to large error bars in the data),
there are some striking features that emerge from the best fit parameters.
For any given $\kappa$, we see that there is a large increase in the star formation efficiency factor ($f_\star$) for redshifts $z > 10$ even though for $\kappa = 4$, $f_\star$ remains nearly constant over $2 \le z \le 10$, whereas for other $\kappa$ values it is approximately constant only over $6 \le z \le 10$.
Between $z = 10$ and $z = 12.5$, $f_\star$ increases by factor of approximately $2$ for models with $\kappa = 1$ and $\kappa = 0.5$.
Even in the $\kappa = 0.25$ model, where the star formation timescale is shorter, there is still a notable increase by a factor of $\sim 1.6$.
This upward trend becomes even steeper when extended to $z = 14$.
If $\kappa$ is held fixed, the star formation efficiency $f_\star$ increases by factors of 3.4, 3.2, and 2.4 for the $\kappa = 1$, 0.5, and 0.25 models, respectively, compared to its value at $z = 10$.
For better demonstration of the evolution of $f_*$ across redshifts, in Figure~\ref{fig:f_star_vs_z} we have plotted the fitted values of $f_*$ and the associated uncertainty for models with different values of $\kappa$.
These results indicate a significant shift in the star formation efficiency within a short time interval. 
This may be unphysical as the time interval between $z=10$ to $z=14$ is only
about 180~Myr.
Thus we favour the models with gradual change in the star formation time scale that has been established in the $z\le 10$ to continue in the $z\ge 10$ universe.
For example, we can see from Table~\ref{table:all} it requires an unphysical $f_\star>1$ for $\kappa=4$ model to fit the luminosity function at $z=14$, thus completely ruled out.
However, models assuming $\kappa=0.5$ at $z=10$ and $\kappa=0.25$ at $z=14$ fit the luminosity function with similar $f_\star=0.069$ and 0.064 with reasonably low chi-square values.
Therefore, we conclude that the observed evolution of the UV luminosity function can be understood as the evolution of the dark matter haloes as expected from $\Lambda$CDM cosmology along with a gradual decrease in the star formation time scale with increasing redshift.

It is already known that the star formation time scale is directly related to the halo’s gas depletion timescale.
Studies indicate that the star formation rate depends non-linearly on the halo’s gas density \citep{Schmidt..1959, Kennicutt..1998}.
Recent studies by \cite{Vallini..2024} further suggest that elevated gas densities in the early universe may drive bursty star formation, leading to significantly shorter gas depletion timescales in the halo as also revealed in our present work.
Several other studies have also shown that bursty star formation at high redshift can account for the observed abundance of bright UV-luminous galaxies at $z \gtrsim 10$ \citep{sun..2023a, sun..2023b, Rojas-Ruiz..2025,kokorev..2025, Munoz..2026}. 
Further support for such a bursty behaviour comes from the discovery of post-starburst quenched galaxies at $z \gtrsim 7$ \citep{Messa..2025, Weibel..2025}, and even at $z \sim 11$ \citep{Harikane..2026}, indicating that galaxies can undergo rapid and intense early star formation followed by efficient quenching.
Taken together, these results qualitatively support our model, in which a decreasing star formation timescale toward higher redshift naturally leads to burstier star formation.
This, in turn, is sufficient to reproduce the observed evolution of the UV luminosity functions over a wide redshift range, $2 \le z \le 14$.

\section{Comparative Analysis with Other Models} \label{other models}

As revealed in the previous section, gradual decrease in the star formation timescale with redshift provides a good explanation to the observed luminosity function upto $z=14$, there are other possible explanations that have been explored by various authors \citep[i.e.][]{Ferrara..2023,Dekel..2023,Shen..2023,Yung..2024}.
Here we are exploring such possibilities in order to break the degeneracy among the various models including ours.

\subsection{Dust Attenuation Models}
First, we examine the impact of dust attenuation on the UV LF, as discussed by \citet{Ferrara..2023}.
They proposed a physical model in which extremely high (super-Eddington) specific star formation rates (sSFRs) drive dust outflows from galaxies, effectively rendering them dust-free.
When extended to higher redshifts, the model predicts a slowdown in UV LF evolution from $z \sim 9$ to $z \sim 11$, driven by reduced dust content and lower UV attenuation in early galaxies.

\begin{figure}
\centering

\includegraphics[width=\columnwidth]{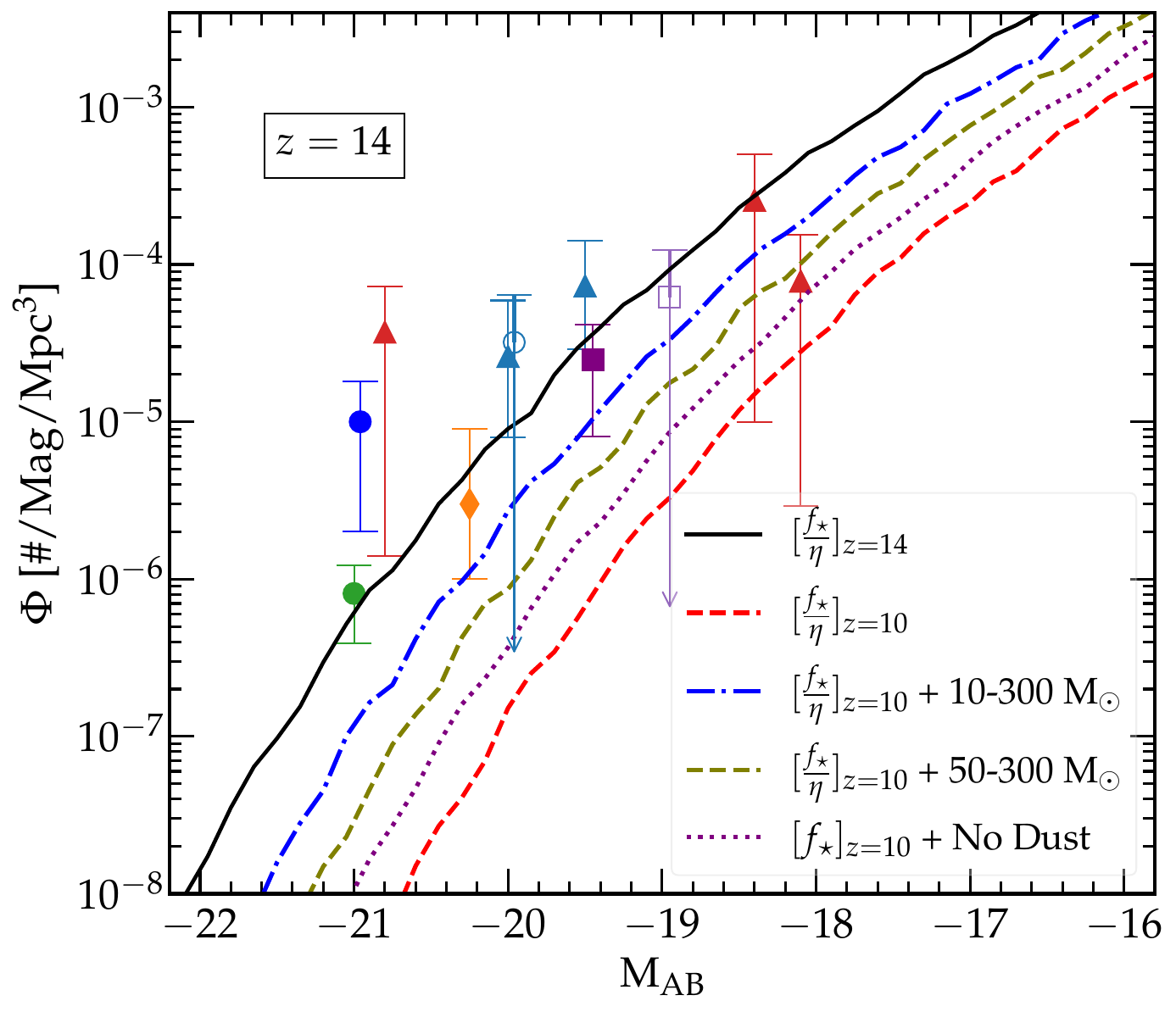}%

\caption[]{Comparison of UV luminosity functions (LFs) at $z=14$ for different models. The solid black line shows the fiducial UV LF at $z=14$ using the best-fit $f_\star/\eta$ parameters for this redshift. The red dashed line represents the UV LF at $z=14$ assuming the best-fit $f_\star/\eta$ values derived for $z=10$, while the blue dash-dotted line incorporates a top-heavy initial mass function (IMF) under the same $z=10$ parameters. The purple dotted line corresponds to the best-fit $f_\star$ at $z=10$ with no dust attenuation ($\eta=1$). In the all above mentioned models we fixed the value $\kappa=1$.}

\label{fig3}
\end{figure}

To understand the effect of dust attenuation, in Figure~\ref{fig3}, we have shown the UV luminosity function as expected from our dust free model at $z=14$ by the purple dotted line.
Here we have used $\kappa=1$ and $f_\star=0.069$ which fit the UV LF at $z=10$ and assumed $\eta=1$ which makes our model dust free. 
It is evident from the figure that the complete absence of dust alone is insufficient to elevate the UV LF to the level required to match the current observational constraints at $z = 14$.
Note that for a better comparison, we have also plotted (red dashed line in figure \ref{fig3}) the expected UV LF at $z=14$ for a model that assumes the $\kappa$, $f_\star$  and $\eta$ do not evolve between $z=10$ and $z=14$ whereas the number density of the dark matter haloes evolves as expected from the $\Lambda$CDM model.
Both models completely under predict the observed UV LF at $z=14$ and hence cannot explain the observation.
This is expected as at $z=10$ the dust attenuation factor $\eta=1.353$ (i.e., $A_\text{FUV}=0.33$) is already low, making a galaxy dimmer by 0.33 magnitude only.
Similar comparison has also been reported by \citet{Finkelstein..2024} that shows no significant evolution in the median far-UV (FUV) color from $z \sim 8$ to $z \sim 11$.

\subsection{Top-Heavy IMF Models}
We further explore how a top-heavy initial mass function (IMF) affects the UV LF of galaxies, particularly at high redshifts.
As discussed by \citet{Harikane..2023a}, \citet{Finkelstein..2023}, and \citet{Finkelstein..2024}, physical conditions in the early universe such as higher cosmic microwave background (CMB) temperature and low gas phase metallicity are conducive to a shift in the characteristic stellar mass.
Under such environments, the typical stellar mass may increase from $\sim 1~\rm M_\odot$ to $\sim10~\rm M_\odot$, as predicted by theoretical studies \citep[i.e.,][]{Larson..1998, Bromm..2002, Tumilson..2006, Steinhardt..2023}, thus producing a top-heavy initial mass function.
In order to investigate the effect of such top-heavy IMF in the UV LF, we explore two possible top-heavy IMF models with the mass range: (i) $10-300~\rm M_\odot$ and (ii) $50-300~\rm M_\odot$, both having a power-law slope of $-2.35$.

In Figure~\ref{fig3}, the blue dash-dotted line and the olive dotted line represent the UV LF at $z \sim 14$ for IMFs with mass ranges of $10-300~\rm M_\odot$ and $50-300~\rm M_\odot$, respectively.
In both of these models, we consider the values of $f_\star=0.069$ and $\kappa=1$ that fit the observed UV LF at $z=10$ with previously assumed Salpeter IMF with mass range $1-100~\rm M_\odot$.
It is clear from the figure that both models underpredict the observed UV LF at $z=14$.
Thus only changing to top-heavy IMF alone cannot explain the non-evolution of the observed UV LF between $z=10$ to $z=14$.
A decrease in $\kappa$ or an increase in $f_\star$ along with a change in IMF is needed to explain the UV LF at $z=14$.  
It is interesting to note that the UV LF for the model with the $50-300~\rm M_\odot$ IMF lies significantly below that of the $10-300~\rm M_\odot$ case.
This suggests that a more massive lower mass cutoff in the IMF requires a larger adjustment in $f_\star$ and/or $\kappa$ compared to a less massive cutoff.
For example, in the $\kappa = 1$ model, the UV LF can be well fitted with $f_\star = 0.094$ ($\chi^2_\nu=1.70$) for an IMF of $10$–$300~\rm M_\odot$, whereas for IMF of $50$–$300~\rm M_\odot$, a higher $f_\star = 0.123$ ($\chi^2_\nu=1.71$) is required.
This may seem counter intuitive as heavier stars are more UV bright.
However, this can be understood by the UV luminosity produced by a single galaxy under the assumption of these two top-heavy IMF. 

\begin{figure}
\centering
\includegraphics[width=\columnwidth]{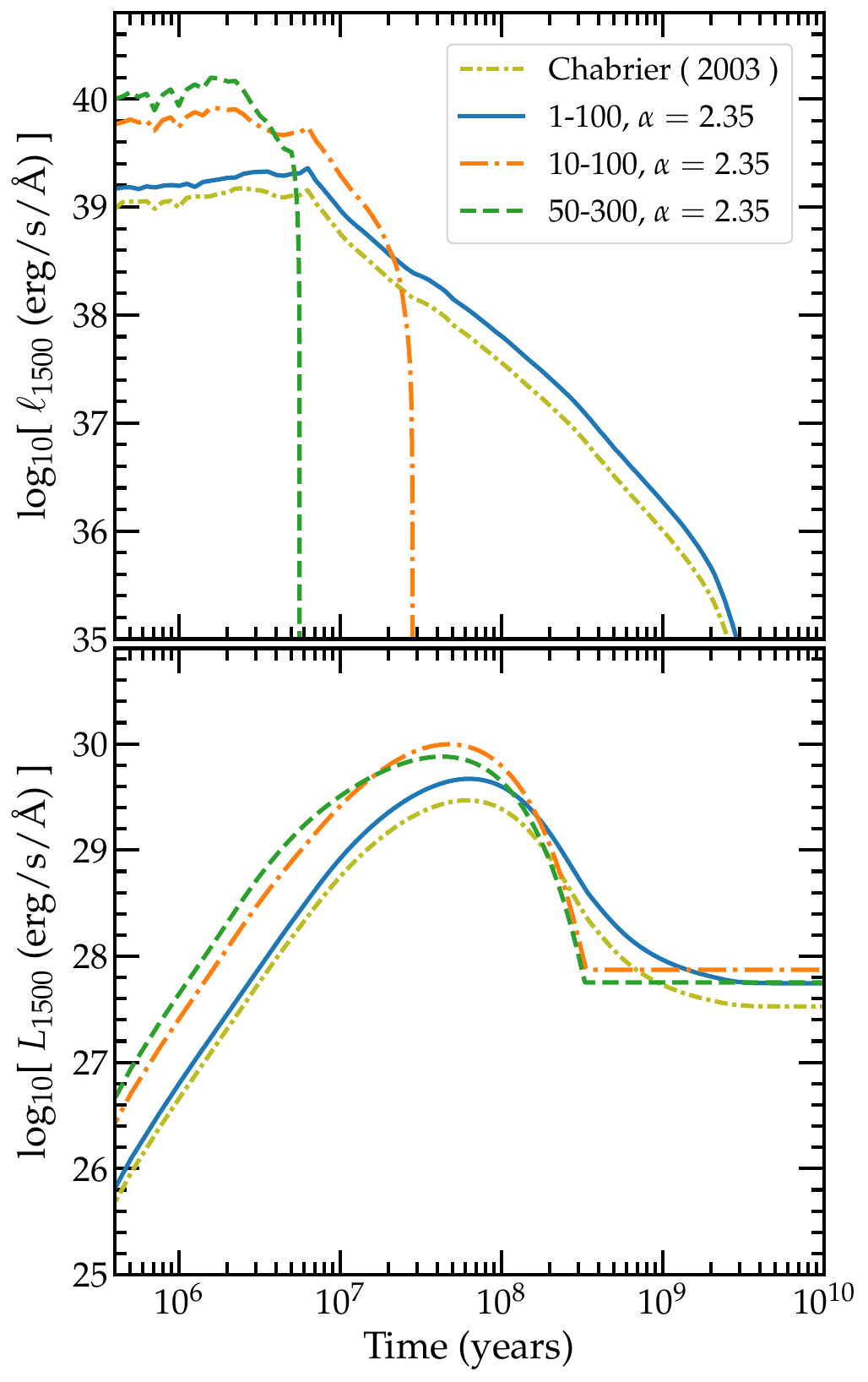}
\caption[]{
\textbf{Upper Panel:} Time evolution of the UV luminosity ($\ell_{1500}$) for a burst of $10^6$~M$_\odot$ star formation with four different initial mass functions: (i) Chabrier IMF \citep{Chabrier..2003}, (ii) Standard Salpeter IMF ($1{-}100~\rm M_\odot$), (iii) Top-Heavy IMF ($10{-}300~\rm M_\odot$), and (iv) More Extreme Top-Heavy IMF ($50{-}300~\rm M_\odot$), all assuming a power-law slope of $-2.35$. \textbf{Lower Panel:} UV luminosity ($L_{1500}$) of a $10^{11}~\rm M_\odot$ halo, formed at $z_c = 15$ with continuous star formation as obtained from equations~\ref{eq:t<t_sn} and \ref{eq:t>t_sn}, shown for the same four IMFs. The flatness in $L_{1500}$ at late times ($\gtrsim 300~\mathrm{Myr}$) arises due to the inclusion of a passive star formation phase in the SFR model (see section \ref{sec-sfr}).
}

\label{fig:fsps_uv1500}
\end{figure}

\par
In the upper panel of Figure \ref{fig:fsps_uv1500}, we have shown the time evolution of the UV luminosity at 1500~\AA~($\ell_{1500}$) for a stellar population with mass of $10^6 \rm M_\odot$ formed in a single instantaneous burst at $t=0$.
We plot $\ell_{1500}$ for four different initial mass functions,  i)~Chabrier IMF \citep{Chabrier..2003}, ii)~Salpeter IMF ($1-100~\rm M_\odot$), iii)~Top-Heavy IMF ($10-300~\rm M_\odot$) and iv)~Extreme Top-Heavy IMF ($50-300~\rm M_\odot$), all assuming a power-law slope of $-2.35$  (i.e. for (ii), (iii) and (iv)).
These models are generated using the Flexible Stellar Population Synthesis (FSPS) code \citep{Conroy..2009}.
The plots show a slight boost in UV output for the Salpeter IMF ($1-100~\rm M_\odot$) compared to the Chabrier IMF, but both follow a similar trend.
In contrast, the top-heavy IMFs ($10-300~\rm M_\odot$ and $50-300~\rm M_\odot$) produce a substantially enhanced UV output compared to the Chabrier and Salpeter IMFs.
This enhancement arises because massive stars contribute more UV light per unit stellar mass than their lower-mass counterparts.
However, due to the rapid evolution and short lifetime of more massive stars, stellar populations with a top-heavy IMF exhibit a steep decline in UV luminosity over time.
This can also be seen in the upper panel of Figure~\ref{fig:fsps_uv1500}.
The $\ell_{1500}$ for the IMF with a mass range of $50-300~\rm M_\odot$ declines at a much earlier time ($\sim5$ Myr) compared to the IMF with a mass range of $10-300~\rm M_\odot$, which declines at a later time ($\sim20$ Myr).
This leads to a lower total UV luminosity under a continuous star formation scenario as can be seen in the lower panel of Figure~\ref{fig:fsps_uv1500}.
The lower panel shows the time evolution of total UV luminosity ($L_{1500}$) under a continuous star formation scenario within a dark matter halo of mass $10^{11}~\rm M_\odot$ that collapses at redshift $z_c = 15$ (calculated using equation~\ref{e3}).
As expected, both top-heavy IMF models initially exhibit significantly higher UV luminosities at early times ($\lesssim 10~\mathrm{Myr}$) due to the dominance of UV-bright, high-mass stars.
In particular, the $50-300~\rm M_\odot$ IMF produces a higher luminosity upto $10^7$ yrs which is the lifetime of a $50~\rm M_\odot$ star.
After which the luminosity for model with $10-300~\rm M_\odot$ becomes higher.
The $10-300~\rm M_\odot$ IMF retains stars of mass in the range $10-50~\rm M_\odot$, thus the UV luminosity remains higher compared to the model with $50-300~\rm M_\odot$ IMF for $t\gtrsim10^7$ yrs.
This results in a lower UV luminosity function as already shown in Figure~\ref{fig3}.
Thus it explains why the model with more massive lower mass cut off in the IMF mass range requires more change in $f_\star$ and/or $\kappa$ compared to that of less massive lower mass cut off to explain the observed UV LF at $z=14$.

\subsection{Stellar mass analysis to break the degeneracy}
\begin{figure}
\centering
\includegraphics[width=0.48\textwidth]{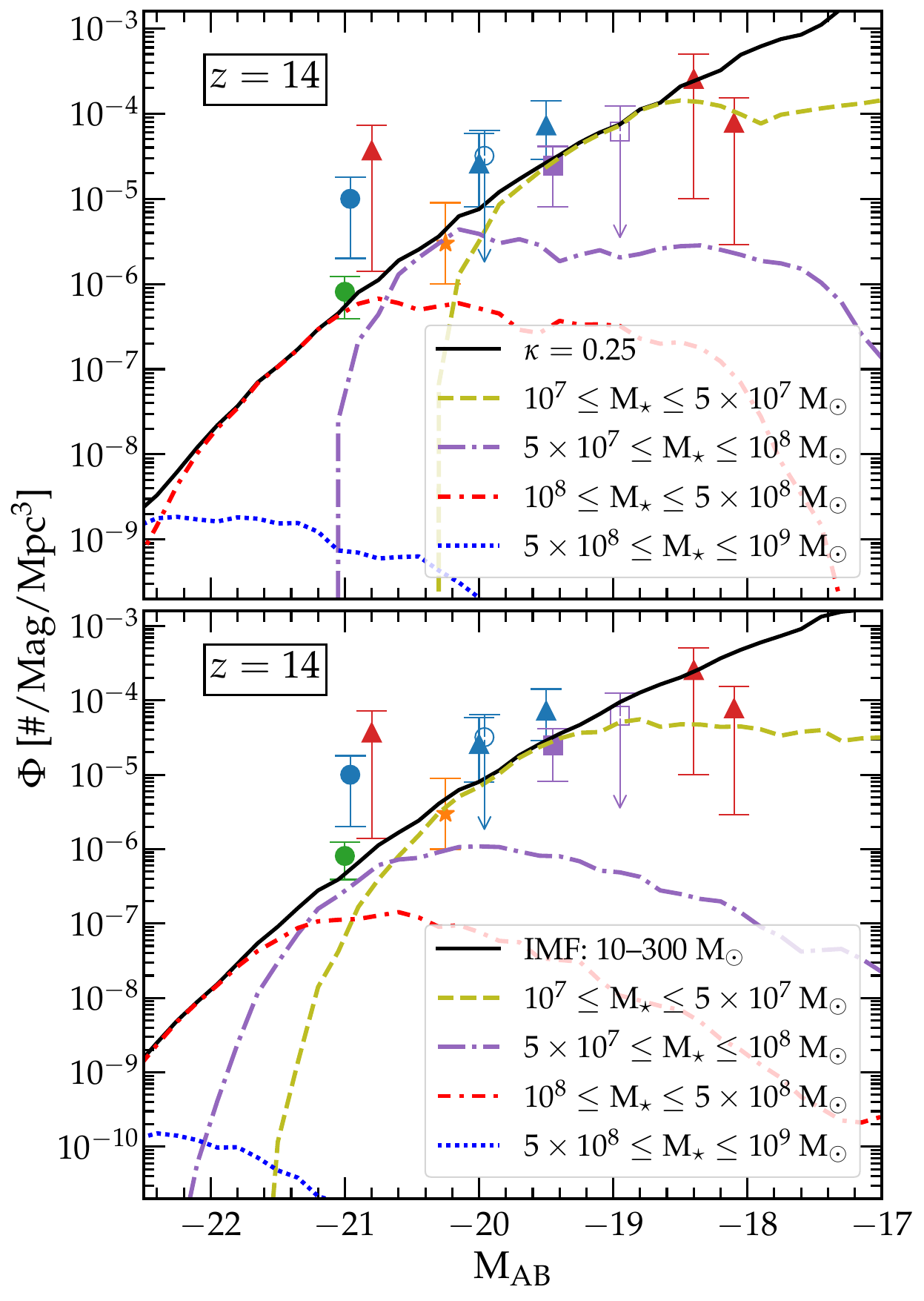}

\caption[]{
Contribution to the UV luminosity function (UV LF) from galaxies in different stellar mass ranges, shown for two models:
i)~$\kappa = 0.25$ with Salpeter IMF ($1-100~\rm M_\odot$) (upper panel), and
ii)~Top-heavy IMF ($10-300~\rm M_\odot$) with $\kappa = 1$ (lower panel).
In both panels, the solid black line represents the best-fit UV LF. The contribution from galaxies in different stellar mass ranges are indicated as follows: dashed olive line for $10^{7}-5\times10^{7}~\rm M_\odot$, long dash-dotted purple line for $5\times10^{7}-10^{8}~\rm M_\odot$, short dashed-dotted red line for $10^{8}-5\times10^{8}~\rm M_\odot$, and dotted blue line for $5\times10^{8}-10^{9}~\rm M_\odot$.
}
\label{fig:mass_range}
\end{figure}

In the previous section we have seen that both bursty star formation model and top-heavy IMF model can potentially explain the observed UV LF at $z=14$.
However the statistical analysis (i.e., $\chi^2_\nu$) alone cannot suggest any preference between them. 
Note that apart from the luminosity functions we also have the estimate of the stellar masses of the detected galaxies using SED fitting techniques.
Here we further explore if such estimates of stellar mass can break the degeneracy between different models. 
To do that, in Figure~\ref{fig:mass_range}, we show the contribution to the UV LF from galaxies in different stellar mass ranges as predicted from our two models:
i)~$\kappa = 0.25$ with Salpeter IMF ($1-100~\rm M_\odot$) (upper panel), and
ii)~Top-Heavy IMF ($10-300~\rm M_\odot$) with $\kappa = 1$ (lower panel).

It is clear from the top panel of the figure that the observed range of UV LF ($-21 \lesssim \text{M}_\text{AB} \lesssim -18$) is primarily contributed by galaxies with stellar masses in the range $10^7-10^8\rm M_\odot$ for the bursty star formation model with $\kappa=0.25$.
Moreover, galaxies with absolute magnitudes $\text{M}_\text{AB} \lesssim -21$ are generally associated with higher stellar masses ($M_\star > 10^8\rm M_\odot$) for this model.
This is in good agreement with the stellar masses derived by \citet{Robertson..2024}, who also estimated the stellar masses of several galaxies with photometric redshifts $z_{\rm{phot}} > 11.5$, primarily falling in the range $7\lesssim \log_{10}(M_\star/\rm M_\odot) \lesssim 8$.
In contrast, the model with the top-heavy IMF shows that the same luminosity range is contributed by galaxies with less stellar mass (bottom panel).
In particular, galaxies with stellar masses $ M_\star > 10^8~\rm M_\odot$ contribute significantly to the UV LF only at magnitudes $\text{M}_\text{AB} \lesssim -21.5$.
Thus, with the help of the stellar masses of the detected galaxies, one can potentially break the degeneracy between these models.

However, the uncertainty in the stellar mass estimates, at present, is as large as $\sim 0.5$~dex.
Further, these estimates depend sensitively on the assumed IMF and star formation timescales.
For example, \citet{Robertson..2024} derived the stellar mass of a spectroscopically confirmed galaxy, JADES-GS-z14-0, at $z_{\text{spec}} \sim 14.32$.
This galaxy has an absolute magnitude of $\text{M}_\text{AB} \sim -21$, and its inferred stellar mass is $\log_{10}( M_\star/\rm M_\odot) = 8.86^{+0.35}_{-0.03}$. They had assumed a Chabrier IMF and a non-parametric star formation history while estimating the stellar mass.
This inferred result should not be compared with model predictions with different IMFs such as top-heavy IMFs.
Therefore, to enable meaningful comparisons, stellar masses should be derived and compared self-consistently assuming the same assumptions about the IMF and other parameters such as star formation history.
To demonstrate the effect of the IMF and star formation history on the derived stellar mass and to highlight the importance of self-consistent analysis, we use \texttt{Prospector} \citep{Johnson..2021}, a fully Bayesian framework designed to extract stellar population parameters from broadband photometry, in the next section.

\subsection{Insights from Prospector} \label{prospector}
We use \texttt{Prospector} to estimate the stellar mass of the spectroscopically confirmed $z_\text{spec} \sim 14.32$ galaxy JADES-GS-z14-0, based on multi-band JADES observations.
We use fluxes in 18 photometric bands (i.e., 5 HST ACS bands + 13 JWST NIRCam bands) given in \cite{Robertson..2024}.
Following the methodology of \citet{Tacchella..2022}, we adopt a delayed–$\tau$ star formation history (SFH), where $\tau$ defines the characteristic timescale.
This star formation model also resembles the model we used in this paper.
This approach enables us to systematically explore how variations in the IMF and SFH influence the inferred stellar mass and age, ensuring a self-consistent analysis.

\begin{table}
\centering
\renewcommand{\arraystretch}{1.4} % Increases row height for better readability
\begin{tabular}{lccc}
\toprule
IMF & $\log_{10}(M_\star)$ & Age (Myr) & $\tau$ (Myr) \\
\midrule
Chabrier (2003) & $8.79^{+0.13}_{-0.11}$ & $17^{+5}_{-3}$ & $0.3^{+0.6}_{-0.2}$ \\
Standard IMF & $8.55^{+0.12}_{-0.11}$ & $17^{+5}_{-3}$ & $0.3^{+0.5}_{-0.2}$ \\
Top-Heavy IMF & $8.24^{+0.15}_{-0.12}$ & $15^{+2}_{-2}$ & $0.3^{+0.3}_{-0.1}$ \\
\bottomrule
\end{tabular}
\caption{Best‐fit stellar masses, ages and the characteristic star formation timescales ($\tau$) for the $z_{\rm spec}\simeq14.32$ galaxy JADES-GS-z14-0 (NIRCam ID 183348), as derived with \texttt{Prospector} \citep{Johnson..2021} using a delayed‐$\tau$ SFH.  We compare three IMFs: Chabrier IMF \citep{Chabrier..2003}, Standard Salpeter IMF ($1$–$100\,\rm M_\odot$, $\alpha=2.35$), Top-Heavy IMF ($10$–$300\,\rm M_\odot$, $\alpha=2.35$), with $\tau$ as a free parameter. The photometric data is taken from \citet{Robertson..2024}.}

\label{tab:stellar_masses}
\end{table}

\begin{figure}
\centering
\includegraphics[width=\columnwidth]{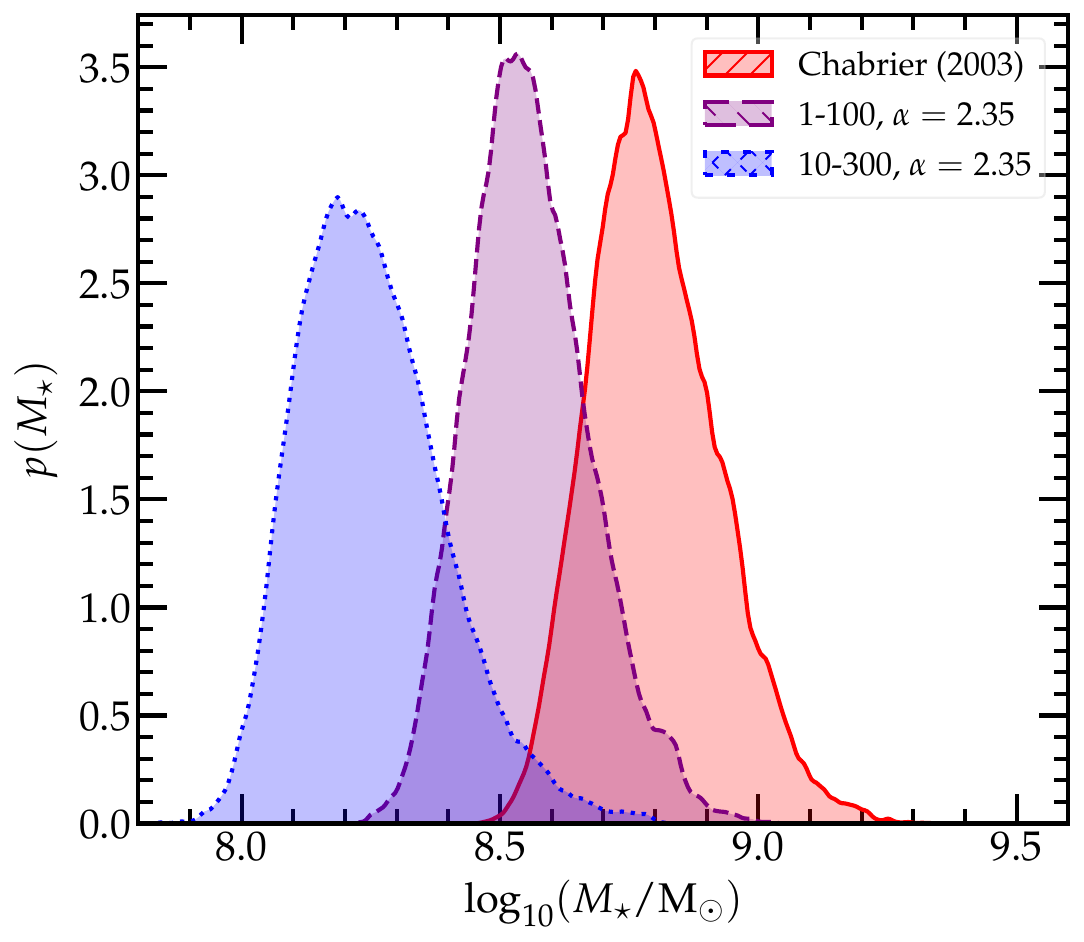}
\caption[]{Posterior distribution of the stellar mass ($M_\star$) for the galaxy JADES-GS-z14-0, derived using the \texttt{Prospector} code \citep{Johnson..2021} under different IMF assumptions. The red, purple, and blue curves correspond to the Chabrier IMF \citep{Chabrier..2003}, Standard Salpeter IMF ($1$–$100\,\rm M_\odot$, $\alpha=2.35$) and Top-Heavy IMF ($10$–$300\,\rm M_\odot$, $\alpha=2.35$) respectively.}
\label{fig:pdf_imf}
\end{figure}

Table~\ref{tab:stellar_masses} summarizes the best‐fit stellar masses, ages, and star formation timescale ($\tau$) values for this galaxy under three IMF assumptions:  
(i) Chabrier IMF \citep{Chabrier..2003},  
(ii) Standard Salpeter‐like IMF (1–100~$\rm M_\odot$, slope = 2.35), and  
(iii) Top‐heavy IMF (10–300~$\rm M_\odot$, slope = 2.35).  
We find that adopting a top‐heavy IMF reduces the inferred stellar mass to $\log(M_\star/\rm M_\odot)=8.24$, compared to $\log(M_\star/\rm M_\odot)=8.79$ with a Chabrier IMF and $\log(M_\star/\rm M_\odot)=8.55$ with our fiducial Salpeter IMF.
In contrast, the estimated ages and $\tau$ values are largely insensitive to the IMF choice.
Thus a top‐heavy IMF shifts the derived stellar mass downward by $\approx0.4$–$0.5$ dex compared to the Chabrier IMF as used in \cite{Robertson..2024}.
The posterior distribution of the derived stellar mass of the galaxy JADES-GS-z14-0, shown in Figure~\ref{fig:pdf_imf}, clearly demonstrates this shift with otherwise similar probability distribution function.

\begin{figure}
\centering
\includegraphics[width=\columnwidth]{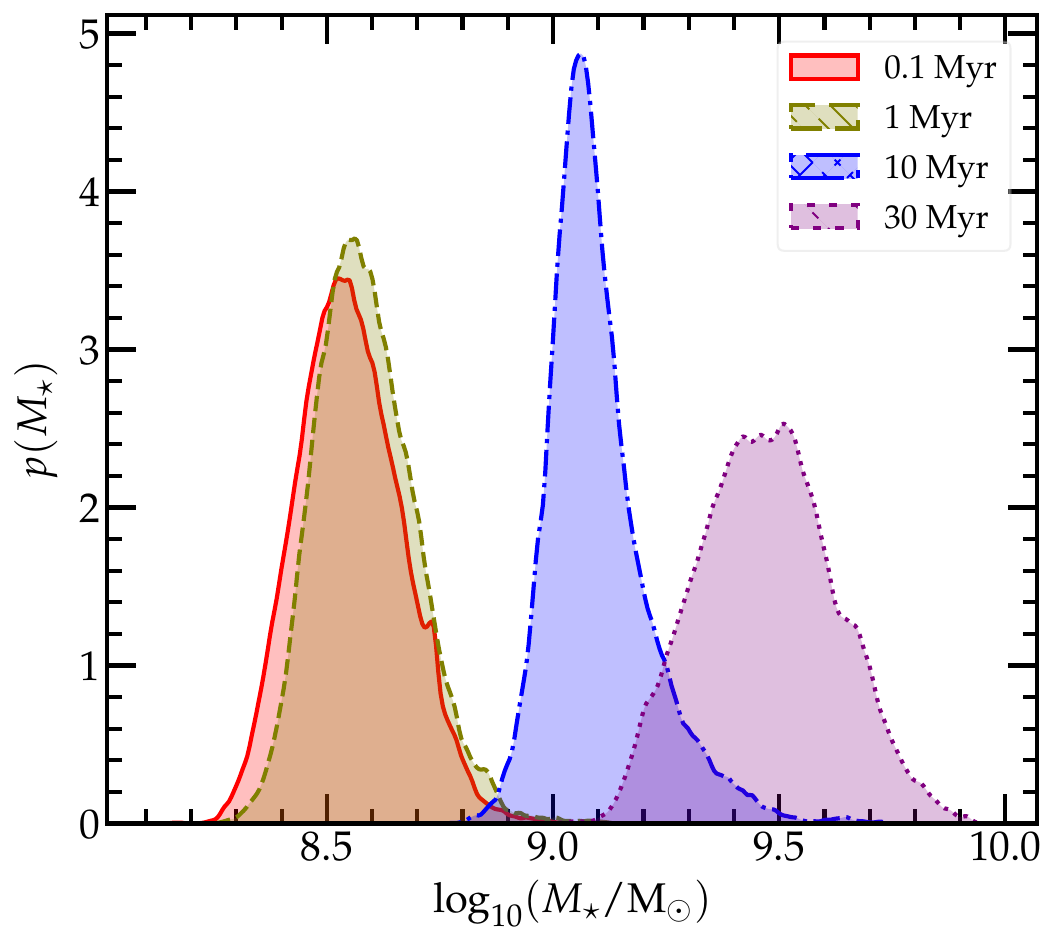}
\caption[]{Posterior distribution of the stellar mass ($M_\star$) for the galaxy JADES-GS-z14-0, derived using the \texttt{Prospector} code \citep{Johnson..2021} by varying the star formation timescale $\tau$, while keeping the IMF fixed. The red, olive, blue, and purple curves correspond to $\tau = 0.1$, 1, 10, and 30~Myr, respectively. As $\tau$ increases, the inferred stellar mass shifts to higher values, indicating the sensitivity of $M_\star$ to assumptions about the star formation history.}
\label{fig:pdf_tau}
\end{figure}

Not only the top-heavy IMF, the values of star formation timescale~($\tau$) also show such a shift as can be seen from Table~\ref{tab:fixed_imf}.
Note that the results presented in Table~\ref{tab:stellar_masses} were obtained by keeping $\tau$ as a free parameter in \texttt{Prospector}.
In Table~\ref{tab:fixed_imf} we show the derived stellar masses resulted from four fixed values of $\tau$ along with the Salpeter IMF.
It is clear from the table that the shortening of $\tau$ from 30 Myr to 0.1 Myr reduces $\log_{10}(M_\star/\rm M_\odot)$ from 9.47 to 8.55 and also lowers the age from 108~Myr to 16~Myr respectively.
Intermediate values of $\tau$ produce correspondingly intermediate masses and ages.
The corresponding posterior distribution of the inferred stellar mass is shown in Figure~\ref{fig:pdf_tau}.
This clearly demonstrates the need for self-consistent comparison of the inferred stellar masses with the appropriate model prediction i.e., one should keep the IMF and $\tau$ value same in the model and in the SED fitting analysis.

\begin{table}
\centering
\renewcommand{\arraystretch}{1.4}
\begin{tabular}{lcc}
\toprule
$\tau$ (Myr) & $\log_{10}(M_\star)$ & Age (Myr) \\
\midrule
0.1 & $8.55^{+0.12}_{-0.11}$ & $16^{+4}_{-2}$ \\
1   & $8.58^{+0.12}_{-0.10}$ & $19^{+4}_{-2}$ \\
10  & $9.08^{+0.12}_{-0.08}$ & $93^{+13}_{-15}$ \\
30  & $9.47^{+0.16}_{-0.15}$ & $108^{+34}_{-40}$ \\
\bottomrule
\end{tabular}
\caption{Same as Table \ref{tab:stellar_masses} but for four fixed characteristic star formation time scale ($\tau$) ($\sim$ 0.1, 1, 10, 30 Myr) assuming the Standard Salpeter IMF ($1$–$100\,\rm M_\odot$, $\alpha=2.35$).}
\label{tab:fixed_imf}
\end{table}

\subsection{Potential AGN Contribution}
In the previous sections we have explored various star formation scenarios in order to understand UV luminosity functions at galaxies $z>10$.
However as pointed out by \cite{Finkelstein..2024} there could be possible contamination of AGN in the sample of galaxies detected in $z>10$.
We therefore explore such possible contamination in this section with a very simple model.

Here we explore a simple model of AGN assuming each halo contains a super massive black hole (SMBH) that accretes the surrounding materials to provide AGN activity in those haloes.
The black hole mass linearly scaled with the halo mass i.e.,  $ M_{{\rm BH}} = \xi M_{{\rm halo}}$, where $\xi$ is the black hole to halo mass ratio.
As shown by \citet{Stuart..2006} this ratio lies between $10^{-4}$-$10^{-3}$ at $z\sim4$ and potentially higher for higher redshift.
Here we consider this entire range while predicting the UV luminosity from the AGN activity.
Further we assume that the AGNs are accreting in sub-Eddingtone regime thus their bolometric luminosity can be modeled as $L_{\rm bol} = \lambda_{\rm Edd} L_{\rm Edd}$, where the Eddington luminosity is given by $ L_{\rm Edd} = 1.26 \times 10^{38} (M_{\rm BH}/\rm M_\odot)$~erg~s$^{-1}$ \citep{Rybicki..1979,Peterson..1997}.
The efficiency factor $\lambda_{\rm Edd}$ is highly uncertain and for our simple model we assume a conservative value of $\lambda_{\rm Edd} = 0.1$ \citep{Vasudevan..2007}. 
Further at 1500~\AA~luminosity due to the AGN activity, $ L^\text{AGN}_{1500}$ is calculated from the bolometric luminosity as $ L^{\rm AGN}_{\rm 1500} = (L_{\rm bol}\times f_\text{UV})/\Delta\nu$ with $\Delta\nu=10^{15}$ Hz appropriate for $\lambda=$ 1500~\AA~\citep{Richards..2006,Runnoe..2012}.
We adopt a bolometric-to-UV conversion factor of $f_{\rm UV} = 0.2$, which is consistent with empirical AGN spectral energy distributions, as shown in \citet{Richards..2006} and \citet{Runnoe..2012}.

Thus, in our simple model the UV luminosity due to the AGN in a halo having mass $M_{\rm halo}$ is given by,
\begin{equation}\label{agn_eqn}
     L^{\rm AGN}_{1500} = 2.52 \times 10^{28} \left(\frac{\lambda_{\rm Edd}}{0.1} \right) \left( \frac{f_{\rm UV}}{0.2} \right) \left(\frac{\xi}{10^{-3}}\right) \left(\frac{M_{\rm halo}}{10^{10}\rm M_\odot}\right)~\mathrm{erg~s^{-1}~Hz^{-1}}.
\end{equation}

\begin{figure}
\centering

\includegraphics[width=\columnwidth]{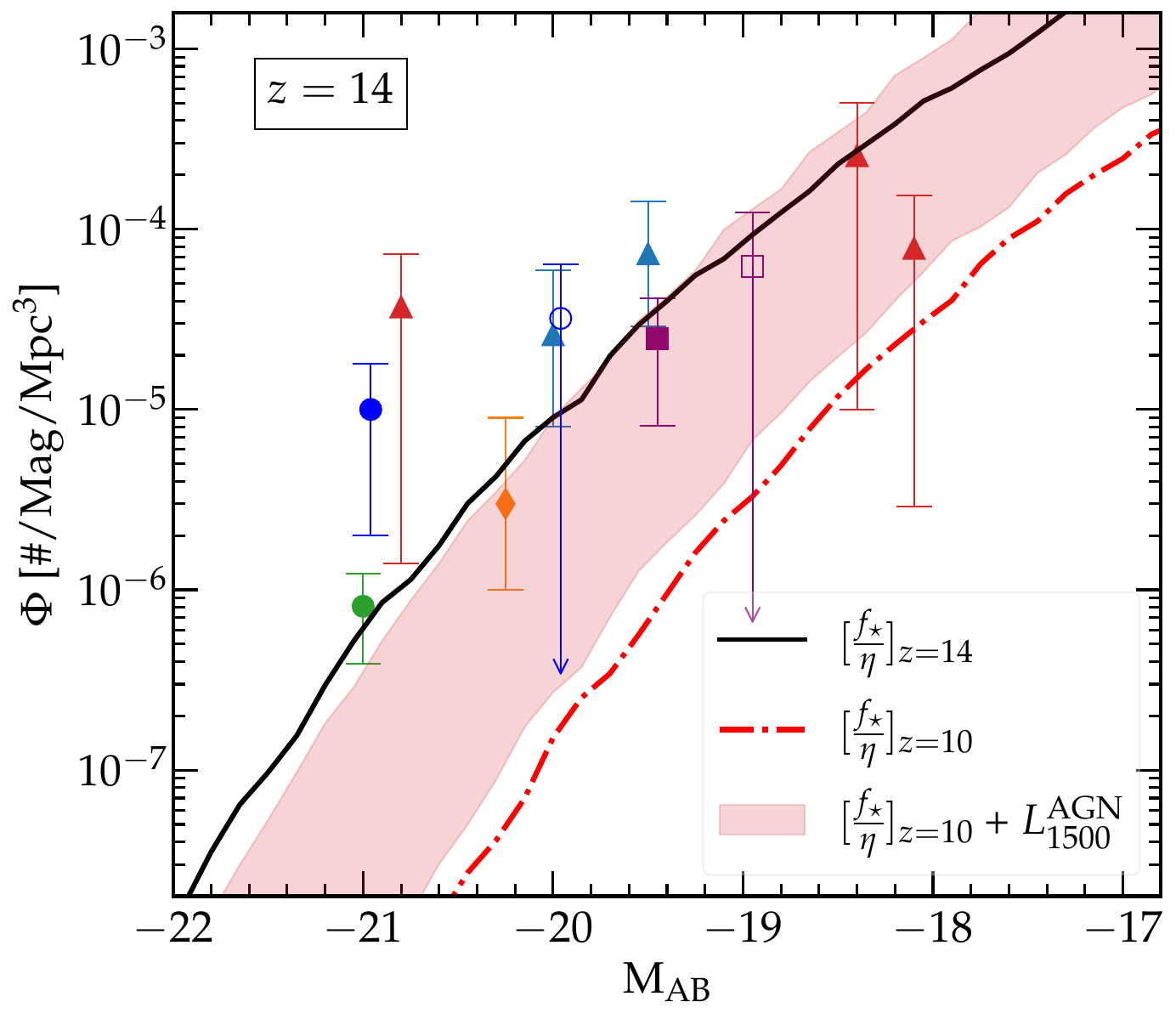}%

\caption[]{Comparison of UV luminosity functions (LFs) at $z=14$ for different models. The solid black line shows the fiducial UV LF at $z=14$ using the best-fit $f_\star/\eta$ parameters for this redshift. The red dashed-dotted line represents the UV LF at $z=14$ assuming the best-fit $f_\star/\eta$ values derived for $z=10$. The red shaded region illustrates the potential AGN contribution to the UV LF at $z=14$ with same best-fit $z=10$ parameters, bounded by two scenarios: a lower limit with a black hole-to-halo mass ratio $M_{\rm{BH}}$/$M_{\rm{halo}} \sim10^{-4}$ and an upper limit with $M_{\rm BH}$/$M_{\rm halo} \sim10^{-3}$.}

\label{fig:AGN}
\end{figure}

The resulting AGN UV luminosity is combined with the stellar UV luminosity to compute the total luminosity function at different redshifts. 
In Figure~\ref{fig:AGN} the red shaded region illustrates the resultant UV LF at $z \sim 14$, when the AGN contribution is added as in equation~\ref{agn_eqn}.
The red dashed line shows the UV LF at $z=14$ when we consider the stellar contribution and the best fit value $f_\star/\eta=0.051$ at $z=10$.
Now the red shaded region shows the enhanced LF when the AGN contribution added with it. 
It is clear from the figure that the moderate parameter value of $\lambda_{\rm Edd}$, $\xi$ and $\rm f_{\rm UV}$ can boost the UV luminosity considerably and can potentially explain the observed UV LF at $z=14$ without changing the stellar parameters.
Thus a detailed spectroscopic study of the detected galaxies is needed in order to confirm the presence of AGNs in those high redshift galaxies.

\section{Discussions and Conclusions}

In this study, we used semi-analytical models of the UV LF to investigate the underlying physical mechanisms responsible for its unexpectedly slow evolution at redshifts $z > 10$, as revealed by recent observations from the JWST.
To better understand the drivers of this transition and the apparent slow evolution of the UV LF at $z > 10$, we systematically explored a range of scenarios using our models.
We found that an evolving star formation timescale plays an important role in explaining the mild redshift evolution of the UV LF at $z > 10$.
We explored the importance of the star formation timescale, low dust attenuation, top-heavy initial mass function, and potential AGN contributions to the observed UV LF at $z > 10$.

Our key findings are summarized below:
\begin{itemize}
    \item In the post-reionization epoch ($z \lesssim 5$), our models favor a longer characteristic star formation timescale with a nearly constant star formation efficiency ($f_\star$) in order to explain the observed UV LFs.

    \item At earlier epochs (i.e., $6 \lesssim z \lesssim 10$), the evolution of the UV LF is better described by a shorter star formation timescale, without requiring a significant change in $f_\star$. This suggests a shift towards burstier star formation mode in the early universe.

    \item Even at $z > 10$, the same trend follows. We find that a gradual shift toward a shorter star formation timescale is favored to explain the observed slow evolution of the UV luminosity function during this epoch without changing the star formation efficiency. Thus our analysis highlights that an evolving star formation timescale can explain the observed UV LF for a wide redshift range of $z=2-14$.

    \item We also show that a dust-free model alone cannot account for the slow evolution of the UV LF at $z > 10$.

    \item Introducing a top-heavy IMF does enhance the UV luminosity of galaxies, but it is unlikely to be the sole driver to explain the observed UV LF evolution. Furthermore, we find that top-heavy IMFs with more massive lower-end cutoffs ($50 - 300~\rm M_\odot$) are less effective than those with lower cutoffs ($10-300~\rm M_\odot$) in reproducing the observed UV LF at $z = 14$.

    \item Our simple model indicates that a moderate AGN activity, with reasonable choices of $\lambda_{\rm Edd}$, $\xi$, and $\rm f_{\rm UV}$, can significantly boost the UV luminosity of high-redshift galaxies. This suggests that the recent observed UV LF at $z \sim 14$ could be explained by possible AGN activity in that epoch without significant change in stellar population, highlighting the importance of future spectroscopic observations to confirm or rule out AGN contamination in these early galaxy samples.
 
    \item We show that the inferred stellar masses of observed galaxies at $z \gtrsim 10$ can help to discriminate between different star formation models.
    However using the Bayesian inference framework \texttt{Prospector}, we also demonstrate that the observed stellar mass should be estimated in a self consistent manner in order to break the degeneracies.

\end{itemize}

One of the major concerns regarding high-redshift galaxy candidates is the reliability of their redshift estimates.
Currently, the number of spectroscopically confirmed sources at these redshifts is limited and heavily biased toward relatively bright galaxies.
Therefore, deep JWST/NIRSpec spectroscopy is critical for robust redshift confirmation and for diagnosing whether the observed emission arises from AGN activity or from exceptionally massive stellar populations.
Further, we now have observations of H$\alpha$ luminosity functions and H$\alpha$/UV ratios of galaxies extending even upto $z\sim 6$ \citep{Sun..2025,Fu..2025,Lin..2026}. Thus, it would be interesting to use our star formation model to understand these observations that we plan to do in the future.

\section*{Acknowledgements}

The authors thank ANRF for financial support through the SERB-SURE grant (SUR/2022/001789).
We thank the anonymous referee for several useful comments that have improved the paper.
SS thanks Presidency University for support through the Faculty Research and Professional Development (FRPDF) Grant.
SS also thanks IUCAA for its support through university associateship programme.

%%%%%%%%%%%%%%%%%%%%%%%%%%%%%%%%%%%%%%%%%%%%%%%%%%
\section*{Data Availability}

The observational ultraviolet (UV) luminosity functions used in this study are taken from the literature cited in the References.
The data generated and analyzed in this work will be made available from the corresponding author upon reasonable request.

\section*{ORCID iDs}

Rupam Sarkar \orcidlink{0009-0000-7016-8085} \href{https://orcid.org/0009-0000-7016-8085}{https://orcid.org/0009-0000-7016-8085}\\
Saumyadip Samui \orcidlink{0000-0002-8052-0967} \href{https://orcid.org/0000-0002-8052-0967}{https://orcid.org/0000-0002-8052-0967}\\



\bibliographystyle{mnras}
\bibliography{ref} % if your bibtex file is called example.bib

\begin{thebibliography}{}
\makeatletter
\relax
\def\mn@urlcharsother{\let\do\@makeother \do\$\do\&\do\#\do\^\do\_\do\%\do\~}
\def\mn@doi{\begingroup\mn@urlcharsother \@ifnextchar [ {\mn@doi@}
  {\mn@doi@[]}}
\def\mn@doi@[#1]#2{\def\@tempa{#1}\ifx\@tempa\@empty \href
  {http://dx.doi.org/#2} {doi:#2}\else \href {http://dx.doi.org/#2} {#1}\fi
  \endgroup}
\def\mn@eprint#1#2{\mn@eprint@#1:#2::\@nil}
\def\mn@eprint@arXiv#1{\href {http://arxiv.org/abs/#1} {{\tt arXiv:#1}}}
\def\mn@eprint@dblp#1{\href {http://dblp.uni-trier.de/rec/bibtex/#1.xml}
  {dblp:#1}}
\def\mn@eprint@#1:#2:#3:#4\@nil{\def\@tempa {#1}\def\@tempb {#2}\def\@tempc
  {#3}\ifx \@tempc \@empty \let \@tempc \@tempb \let \@tempb \@tempa \fi \ifx
  \@tempb \@empty \def\@tempb {arXiv}\fi \@ifundefined
  {mn@eprint@\@tempb}{\@tempb:\@tempc}{\expandafter \expandafter \csname
  mn@eprint@\@tempb\endcsname \expandafter{\@tempc}}}

\bibitem[\protect\citeauthoryear{{Adams} et~al.,}{{Adams}
  et~al.}{2024}]{Adams..2024}
{Adams} N.~J.,  et~al., 2024, \mn@doi [\apj] {10.3847/1538-4357/ad2a7b}, \href
  {https://ui.adsabs.harvard.edu/abs/2024ApJ...965..169A} {965, 169}

\bibitem[\protect\citeauthoryear{{Andrews} \& {Martini}}{{Andrews} \&
  {Martini}}{2013}]{Andrews..2013}
{Andrews} B.~H.,  {Martini} P.,  2013, \mn@doi [\apj]
  {10.1088/0004-637X/765/2/140}, \href
  {https://ui.adsabs.harvard.edu/abs/2013ApJ...765..140A} {765, 140}

\bibitem[\protect\citeauthoryear{{Arrabal Haro} et~al.,}{{Arrabal Haro}
  et~al.}{2023}]{Arrabal..2023}
{Arrabal Haro} P.,  et~al., 2023, \mn@doi [\nat] {10.1038/s41586-023-06521-7},
  \href {https://ui.adsabs.harvard.edu/abs/2023Natur.622..707A} {622, 707}

\bibitem[\protect\citeauthoryear{{Asplund}, {Grevesse}, {Sauval}  \&
  {Scott}}{{Asplund} et~al.}{2009}]{Asplund..2009}
{Asplund} M.,  {Grevesse} N.,  {Sauval} A.~J.,   {Scott} P.,  2009, \mn@doi
  [\araa] {10.1146/annurev.astro.46.060407.145222}, \href
  {https://ui.adsabs.harvard.edu/abs/2009ARA&A..47..481A} {47, 481}

\bibitem[\protect\citeauthoryear{Atek, Richard, Kneib  \& Schaerer}{Atek
  et~al.}{2018}]{Atek..2018}
Atek H.,  Richard J.,  Kneib J.-P.,   Schaerer D.,  2018, \mn@doi [\mnras]
  {10.1093/mnras/sty1820}, 479, 5184

\bibitem[\protect\citeauthoryear{{Bagley} et~al.,}{{Bagley}
  et~al.}{2024}]{Bagley..2024}
{Bagley} M.~B.,  et~al., 2024, \mn@doi [\apj] {10.3847/1538-4357/ad09dc}, \href
  {https://ui.adsabs.harvard.edu/abs/2024ApJ...961..209B} {961, 209}

\bibitem[\protect\citeauthoryear{{Barkana} \& {Loeb}}{{Barkana} \&
  {Loeb}}{2001}]{Barkana_Loeb..2001}
{Barkana} R.,  {Loeb} A.,  2001, \mn@doi [\physrep]
  {10.1016/S0370-1573(01)00019-9}, \href
  {https://ui.adsabs.harvard.edu/abs/2001PhR...349..125B} {349, 125}

\bibitem[\protect\citeauthoryear{{Benson}, {Lacey}, {Baugh}, {Cole}  \&
  {Frenk}}{{Benson} et~al.}{2002}]{Benson..2002}
{Benson} A.~J.,  {Lacey} C.~G.,  {Baugh} C.~M.,  {Cole} S.,   {Frenk} C.~S.,
  2002, \mn@doi [\mnras] {10.1046/j.1365-8711.2002.05387.x}, \href
  {https://ui.adsabs.harvard.edu/abs/2002MNRAS.333..156B} {333, 156}

\bibitem[\protect\citeauthoryear{{Best}, {Kaiser}, {Heckman}  \&
  {Kauffmann}}{{Best} et~al.}{2006}]{Best..2006}
{Best} P.~N.,  {Kaiser} C.~R.,  {Heckman} T.~M.,   {Kauffmann} G.,  2006,
  \mn@doi [\mnras] {10.1111/j.1745-3933.2006.00159.x}, \href
  {https://ui.adsabs.harvard.edu/abs/2006MNRAS.368L..67B} {368, L67}

\bibitem[\protect\citeauthoryear{{Bouwens}, {Illingworth}, {Franx}  \&
  {Ford}}{{Bouwens} et~al.}{2007}]{Bouwens..2007}
{Bouwens} R.~J.,  {Illingworth} G.~D.,  {Franx} M.,   {Ford} H.,  2007, \mn@doi
  [\apj] {10.1086/521811}, \href
  {https://ui.adsabs.harvard.edu/abs/2007ApJ...670..928B} {670, 928}

\bibitem[\protect\citeauthoryear{{Bouwens} et~al.,}{{Bouwens}
  et~al.}{2012}]{Bouwens..2012}
{Bouwens} R.~J.,  et~al., 2012, \mn@doi [\apj] {10.1088/0004-637X/754/2/83},
  \href {https://ui.adsabs.harvard.edu/abs/2012ApJ...754...83B} {754, 83}

\bibitem[\protect\citeauthoryear{{Bouwens} et~al.,}{{Bouwens}
  et~al.}{2015}]{Bouwens..2015}
{Bouwens} R.~J.,  et~al., 2015, \mn@doi [\apj] {10.1088/0004-637X/803/1/34},
  \href {https://ui.adsabs.harvard.edu/abs/2015ApJ...803...34B} {803, 34}

\bibitem[\protect\citeauthoryear{Bouwens, Oesch, Illingworth, Ellis  \&
  Stefanon}{Bouwens et~al.}{2017}]{Bouwens..2017}
Bouwens R.~J.,  Oesch P.~A.,  Illingworth G.~D.,  Ellis R.~S.,   Stefanon M.,
  2017, \mn@doi [\apj] {10.3847/1538-4357/aa70a4}, 843, 129

\bibitem[\protect\citeauthoryear{{Bouwens}, {Stefanon}, {Oesch}, {Illingworth},
  {Nanayakkara}, {Roberts-Borsani}, {Labb{\'e}}  \& {Smit}}{{Bouwens}
  et~al.}{2019}]{Bouwens..2019}
{Bouwens} R.~J.,  {Stefanon} M.,  {Oesch} P.~A.,  {Illingworth} G.~D.,
  {Nanayakkara} T.,  {Roberts-Borsani} G.,  {Labb{\'e}} I.,   {Smit} R.,  2019,
  \mn@doi [\apj] {10.3847/1538-4357/ab24c5}, \href
  {https://ui.adsabs.harvard.edu/abs/2019ApJ...880...25B} {880, 25}

\bibitem[\protect\citeauthoryear{{Bouwens} et~al.,}{{Bouwens}
  et~al.}{2021}]{Bouwens..2021}
{Bouwens} R.~J.,  et~al., 2021, \mn@doi [\aj] {10.3847/1538-3881/abf83e}, \href
  {https://ui.adsabs.harvard.edu/abs/2021AJ....162...47B} {162, 47}

\bibitem[\protect\citeauthoryear{Bouwens, Illingworth, van Dokkum, Oesch,
  Stefanon  \& Ribeiro}{Bouwens et~al.}{2022}]{Bouwens..2022a}
Bouwens R.~J.,  Illingworth G.~D.,  van Dokkum P.~G.,  Oesch P.~A.,  Stefanon
  M.,   Ribeiro B.,  2022, \mn@doi [\apj] {10.3847/1538-4357/ac4791}, 927, 81

\bibitem[\protect\citeauthoryear{{Bouwens}, {Illingworth}, {Oesch}, {Stefanon},
  {Naidu}, {van Leeuwen}  \& {Magee}}{{Bouwens} et~al.}{2023}]{Bouwens..2023}
{Bouwens} R.,  {Illingworth} G.,  {Oesch} P.,  {Stefanon} M.,  {Naidu} R.,
  {van Leeuwen} I.,   {Magee} D.,  2023, \mn@doi [\mnras]
  {10.1093/mnras/stad1014}, \href
  {https://ui.adsabs.harvard.edu/abs/2023MNRAS.523.1009B} {523, 1009}

\bibitem[\protect\citeauthoryear{{Bower}, {Benson}, {Malbon}, {Helly}, {Frenk},
  {Baugh}, {Cole}  \& {Lacey}}{{Bower} et~al.}{2006}]{Bower..2006}
{Bower} R.~G.,  {Benson} A.~J.,  {Malbon} R.,  {Helly} J.~C.,  {Frenk} C.~S.,
  {Baugh} C.~M.,  {Cole} S.,   {Lacey} C.~G.,  2006, \mn@doi [\mnras]
  {10.1111/j.1365-2966.2006.10519.x}, \href
  {https://ui.adsabs.harvard.edu/abs/2006MNRAS.370..645B} {370, 645}

\bibitem[\protect\citeauthoryear{{Bowler} et~al.,}{{Bowler}
  et~al.}{2015}]{Bowler..2015}
{Bowler} R.~A.~A.,  et~al., 2015, \mn@doi [\mnras] {10.1093/mnras/stv1403},
  \href {https://ui.adsabs.harvard.edu/abs/2015MNRAS.452.1817B} {452, 1817}

\bibitem[\protect\citeauthoryear{{Bowler}, {Jarvis}, {Dunlop}, {McLure},
  {McLeod}, {Adams}, {Milvang-Jensen}  \& {McCracken}}{{Bowler}
  et~al.}{2020}]{Bowler..2020}
{Bowler} R.~A.~A.,  {Jarvis} M.~J.,  {Dunlop} J.~S.,  {McLure} R.~J.,  {McLeod}
  D.~J.,  {Adams} N.~J.,  {Milvang-Jensen} B.,   {McCracken} H.~J.,  2020,
  \mn@doi [\mnras] {10.1093/mnras/staa313}, \href
  {https://ui.adsabs.harvard.edu/abs/2020MNRAS.493.2059B} {493, 2059}

\bibitem[\protect\citeauthoryear{{Boylan-Kolchin}}{{Boylan-Kolchin}}{2023}]{Boylan..2023}
{Boylan-Kolchin} M.,  2023, \mn@doi [Nature Astronomy]
  {10.1038/s41550-023-01937-7}, \href
  {https://ui.adsabs.harvard.edu/abs/2023NatAs...7..731B} {7, 731}

\bibitem[\protect\citeauthoryear{{Bromm} \& {Loeb}}{{Bromm} \&
  {Loeb}}{2002}]{Bromm-loeb..2002}
{Bromm} V.,  {Loeb} A.,  2002, \mn@doi [\apj] {10.1086/341189}, \href
  {https://ui.adsabs.harvard.edu/abs/2002ApJ...575..111B} {575, 111}

\bibitem[\protect\citeauthoryear{{Bromm}, {Coppi}  \& {Larson}}{{Bromm}
  et~al.}{2002}]{Bromm..2002}
{Bromm} V.,  {Coppi} P.~S.,   {Larson} R.~B.,  2002, \mn@doi [\apj]
  {10.1086/323947}, \href
  {https://ui.adsabs.harvard.edu/abs/2002ApJ...564...23B} {564, 23}

\bibitem[\protect\citeauthoryear{{Bunker} et~al.,}{{Bunker}
  et~al.}{2024}]{Bunker..2024}
{Bunker} A.~J.,  et~al., 2024, \mn@doi [\aap] {10.1051/0004-6361/202347094},
  \href {https://ui.adsabs.harvard.edu/abs/2024A&A...690A.288B} {690, A288}

\bibitem[\protect\citeauthoryear{{Burgarella} et~al.,}{{Burgarella}
  et~al.}{2013}]{Burgarella..2013}
{Burgarella} D.,  et~al., 2013, \mn@doi [\aap] {10.1051/0004-6361/201321651},
  \href {https://ui.adsabs.harvard.edu/abs/2013A&A...554A..70B} {554, A70}

\bibitem[\protect\citeauthoryear{{Carniani} et~al.,}{{Carniani}
  et~al.}{2024}]{Carniani..2024}
{Carniani} S.,  et~al., 2024, \mn@doi [\nat] {10.1038/s41586-024-07860-9},
  \href {https://ui.adsabs.harvard.edu/abs/2024Natur.633..318C} {633, 318}

\bibitem[\protect\citeauthoryear{{Casey} et~al.,}{{Casey}
  et~al.}{2024}]{Casey..2024}
{Casey} C.~M.,  et~al., 2024, \mn@doi [\apj] {10.3847/1538-4357/ad2075}, \href
  {https://ui.adsabs.harvard.edu/abs/2024ApJ...965...98C} {965, 98}

\bibitem[\protect\citeauthoryear{Castellano et~al.,}{Castellano
  et~al.}{2022}]{Castellano..2022}
Castellano M.,  et~al., 2022, \mn@doi [\apjl] {10.3847/2041-8213/ac94d0}, 938,
  L15

\bibitem[\protect\citeauthoryear{{Chabrier}}{{Chabrier}}{2003}]{Chabrier..2003}
{Chabrier} G.,  2003, \mn@doi [\pasp] {10.1086/376392}, \href
  {https://ui.adsabs.harvard.edu/abs/2003PASP..115..763C} {115, 763}

\bibitem[\protect\citeauthoryear{{Conroy} \& {Gunn}}{{Conroy} \&
  {Gunn}}{2010}]{Conroy..2010}
{Conroy} C.,  {Gunn} J.~E.,  2010, \mn@doi [\apj]
  {10.1088/0004-637X/712/2/833}, \href
  {https://ui.adsabs.harvard.edu/abs/2010ApJ...712..833C} {712, 833}

\bibitem[\protect\citeauthoryear{{Conroy}, {Gunn}  \& {White}}{{Conroy}
  et~al.}{2009}]{Conroy..2009}
{Conroy} C.,  {Gunn} J.~E.,   {White} M.,  2009, \mn@doi [\apj]
  {10.1088/0004-637X/699/1/486}, \href
  {https://ui.adsabs.harvard.edu/abs/2009ApJ...699..486C} {699, 486}

\bibitem[\protect\citeauthoryear{{Curti}, {Mannucci}, {Cresci}  \&
  {Maiolino}}{{Curti} et~al.}{2020}]{Curti..2020}
{Curti} M.,  {Mannucci} F.,  {Cresci} G.,   {Maiolino} R.,  2020, \mn@doi
  [\mnras] {10.1093/mnras/stz2910}, \href
  {https://ui.adsabs.harvard.edu/abs/2020MNRAS.491..944C} {491, 944}

\bibitem[\protect\citeauthoryear{{Curtis-Lake} et~al.,}{{Curtis-Lake}
  et~al.}{2023}]{Curtis..2023}
{Curtis-Lake} E.,  et~al., 2023, \mn@doi [Nature Astronomy]
  {10.1038/s41550-023-01918-w}, \href
  {https://ui.adsabs.harvard.edu/abs/2023NatAs...7..622C} {7, 622}

\bibitem[\protect\citeauthoryear{{Dekel}, {Sarkar}, {Birnboim}, {Mandelker}  \&
  {Li}}{{Dekel} et~al.}{2023}]{Dekel..2023}
{Dekel} A.,  {Sarkar} K.~C.,  {Birnboim} Y.,  {Mandelker} N.,   {Li} Z.,  2023,
  \mn@doi [\mnras] {10.1093/mnras/stad1557}, \href
  {https://ui.adsabs.harvard.edu/abs/2023MNRAS.523.3201D} {523, 3201}

\bibitem[\protect\citeauthoryear{{Dijkstra}, {Haiman}, {Rees}  \&
  {Weinberg}}{{Dijkstra} et~al.}{2004}]{Dijkstra..2004}
{Dijkstra} M.,  {Haiman} Z.,  {Rees} M.~J.,   {Weinberg} D.~H.,  2004, \mn@doi
  [\apj] {10.1086/380603}, \href
  {https://ui.adsabs.harvard.edu/abs/2004ApJ...601..666D} {601, 666}

\bibitem[\protect\citeauthoryear{{Donnan} et~al.,}{{Donnan}
  et~al.}{2023a}]{Donnan..2023a}
{Donnan} C.~T.,  et~al., 2023a, \mn@doi [\mnras] {10.1093/mnras/stac3472},
  \href {https://ui.adsabs.harvard.edu/abs/2023MNRAS.518.6011D} {518, 6011}

\bibitem[\protect\citeauthoryear{{Donnan}, {McLeod}, {McLure}, {Dunlop},
  {Carnall}, {Cullen}  \& {Magee}}{{Donnan} et~al.}{2023b}]{Donnan..2023b}
{Donnan} C.~T.,  {McLeod} D.~J.,  {McLure} R.~J.,  {Dunlop} J.~S.,  {Carnall}
  A.~C.,  {Cullen} F.,   {Magee} D.,  2023b, \mn@doi [\mnras]
  {10.1093/mnras/stad471}, \href
  {https://ui.adsabs.harvard.edu/abs/2023MNRAS.520.4554D} {520, 4554}

\bibitem[\protect\citeauthoryear{{Donnan} et~al.,}{{Donnan}
  et~al.}{2024}]{Donnan..2024}
{Donnan} C.~T.,  et~al., 2024, \mn@doi [\mnras] {10.1093/mnras/stae2037}, \href
  {https://ui.adsabs.harvard.edu/abs/2024MNRAS.533.3222D} {533, 3222}

\bibitem[\protect\citeauthoryear{{Dotter}}{{Dotter}}{2016}]{Dotter..2016}
{Dotter} A.,  2016, \mn@doi [\apjs] {10.3847/0067-0049/222/1/8}, \href
  {https://ui.adsabs.harvard.edu/abs/2016ApJS..222....8D} {222, 8}

\bibitem[\protect\citeauthoryear{Ferrara, Pallottini  \& Dayal}{Ferrara
  et~al.}{2023}]{Ferrara..2023}
Ferrara A.,  Pallottini A.,   Dayal P.,  2023, \mn@doi [\mnras]
  {10.1093/mnras/stad1095}, 522, 3986

\bibitem[\protect\citeauthoryear{{Finkelstein} et~al.,}{{Finkelstein}
  et~al.}{2015}]{Finkelstein..2015}
{Finkelstein} S.~L.,  et~al., 2015, \mn@doi [\apj]
  {10.1088/0004-637X/810/1/71}, \href
  {https://ui.adsabs.harvard.edu/abs/2015ApJ...810...71F} {810, 71}

\bibitem[\protect\citeauthoryear{{Finkelstein} et~al.,}{{Finkelstein}
  et~al.}{2022a}]{Finkelstein..2022a}
{Finkelstein} S.~L.,  et~al., 2022a, \mn@doi [\apj] {10.3847/1538-4357/ac3aed},
  \href {https://ui.adsabs.harvard.edu/abs/2022ApJ...928...52F} {928, 52}

\bibitem[\protect\citeauthoryear{{Finkelstein} et~al.,}{{Finkelstein}
  et~al.}{2022b}]{finklestein..2022}
{Finkelstein} S.~L.,  et~al., 2022b, \mn@doi [\apj] {10.3847/1538-4357/ac3aed},
  \href {https://ui.adsabs.harvard.edu/abs/2022ApJ...928...52F} {928, 52}

\bibitem[\protect\citeauthoryear{{Finkelstein} et~al.,}{{Finkelstein}
  et~al.}{2023}]{Finkelstein..2023}
{Finkelstein} S.~L.,  et~al., 2023, \mn@doi [\apjl] {10.3847/2041-8213/acade4},
  \href {https://ui.adsabs.harvard.edu/abs/2023ApJ...946L..13F} {946, L13}

\bibitem[\protect\citeauthoryear{{Finkelstein} et~al.,}{{Finkelstein}
  et~al.}{2024}]{Finkelstein..2024}
{Finkelstein} S.~L.,  et~al., 2024, \mn@doi [\apjl] {10.3847/2041-8213/ad4495},
  \href {https://ui.adsabs.harvard.edu/abs/2024ApJ...969L...2F} {969, L2}

\bibitem[\protect\citeauthoryear{{Fu} et~al.,}{{Fu} et~al.}{2025}]{Fu..2025}
{Fu} S.,  et~al., 2025, \mn@doi [\apj] {10.3847/1538-4357/adddb1}, \href
  {https://ui.adsabs.harvard.edu/abs/2025ApJ...987..186F} {987, 186}

\bibitem[\protect\citeauthoryear{{Fujimoto} et~al.,}{{Fujimoto}
  et~al.}{2024}]{Fujimoto..2024}
{Fujimoto} S.,  et~al., 2024, \mn@doi [\apj] {10.3847/1538-4357/ad9027}, \href
  {https://ui.adsabs.harvard.edu/abs/2024ApJ...977..250F} {977, 250}

\bibitem[\protect\citeauthoryear{{Harikane} et~al.,}{{Harikane}
  et~al.}{2023a}]{Harikane..2023a}
{Harikane} Y.,  et~al., 2023a, \mn@doi [\apjs] {10.3847/1538-4365/acaaa9},
  \href {https://ui.adsabs.harvard.edu/abs/2023ApJS..265....5H} {265, 5}

\bibitem[\protect\citeauthoryear{{Harikane} et~al.,}{{Harikane}
  et~al.}{2023b}]{Harikane..2023b}
{Harikane} Y.,  et~al., 2023b, \mn@doi [\apj] {10.3847/1538-4357/ad029e}, \href
  {https://ui.adsabs.harvard.edu/abs/2023ApJ...959...39H} {959, 39}

\bibitem[\protect\citeauthoryear{{Harikane} et~al.,}{{Harikane}
  et~al.}{2025}]{Harikane..2025}
{Harikane} Y.,  et~al., 2025, \mn@doi [\apj] {10.3847/1538-4357/ad9b2c}, \href
  {https://ui.adsabs.harvard.edu/abs/2025ApJ...980..138H} {980, 138}

\bibitem[\protect\citeauthoryear{{Harikane} et~al.,}{{Harikane}
  et~al.}{2026}]{Harikane..2026}
{Harikane} Y.,  et~al., 2026, \mn@doi [arXiv e-prints]
  {10.48550/arXiv.2601.21833}, \href
  {https://ui.adsabs.harvard.edu/abs/2026arXiv260121833H} {p. arXiv:2601.21833}

\bibitem[\protect\citeauthoryear{{Ishigaki}, {Kawamata}, {Ouchi}, {Oguri},
  {Shimasaku}  \& {Ono}}{{Ishigaki} et~al.}{2015}]{BoRG..2015}
{Ishigaki} M.,  {Kawamata} R.,  {Ouchi} M.,  {Oguri} M.,  {Shimasaku} K.,
  {Ono} Y.,  2015, \mn@doi [\apj] {10.1088/0004-637X/799/1/12}, \href
  {https://ui.adsabs.harvard.edu/abs/2015ApJ...799...12I} {799, 12}

\bibitem[\protect\citeauthoryear{{Johnson}, {Leja}, {Conroy}  \&
  {Speagle}}{{Johnson} et~al.}{2021}]{Johnson..2021}
{Johnson} B.~D.,  {Leja} J.,  {Conroy} C.,   {Speagle} J.~S.,  2021, \mn@doi
  [\apjs] {10.3847/1538-4365/abef67}, \href
  {https://ui.adsabs.harvard.edu/abs/2021ApJS..254...22J} {254, 22}

\bibitem[\protect\citeauthoryear{Kannan et~al.,}{Kannan
  et~al.}{2023}]{MillenniumTNG}
Kannan R.,  et~al., 2023, \mn@doi [Monthly Notices of the Royal Astronomical
  Society] {10.1093/mnras/stac3743}, 524, 2594

\bibitem[\protect\citeauthoryear{{Kennicutt}}{{Kennicutt}}{1998}]{Kennicutt..1998}
{Kennicutt} Jr. R.~C.,  1998, \mn@doi [\apj] {10.1086/305588}, \href
  {https://ui.adsabs.harvard.edu/abs/1998ApJ...498..541K} {498, 541}

\bibitem[\protect\citeauthoryear{{Khaire} \& {Srianand}}{{Khaire} \&
  {Srianand}}{2015}]{Khaire..2015}
{Khaire} V.,  {Srianand} R.,  2015, \mn@doi [\apj]
  {10.1088/0004-637X/805/1/33}, \href
  {https://ui.adsabs.harvard.edu/abs/2015ApJ...805...33K} {805, 33}

\bibitem[\protect\citeauthoryear{{Kokorev} et~al.,}{{Kokorev}
  et~al.}{2025}]{kokorev..2025}
{Kokorev} V.,  et~al., 2025, \mn@doi [\apjl] {10.3847/2041-8213/ade8f5}, \href
  {https://ui.adsabs.harvard.edu/abs/2025ApJ...988L..10K} {988, L10}

\bibitem[\protect\citeauthoryear{{Larson}}{{Larson}}{1998}]{Larson..1998}
{Larson} R.~B.,  1998, \mn@doi [\mnras] {10.1046/j.1365-8711.1998.02045.x},
  \href {https://ui.adsabs.harvard.edu/abs/1998MNRAS.301..569L} {301, 569}

\bibitem[\protect\citeauthoryear{{Leung} et~al.,}{{Leung}
  et~al.}{2023}]{Leung..2023}
{Leung} G. C.~K.,  et~al., 2023, \mn@doi [\apjl] {10.3847/2041-8213/acf365},
  \href {https://ui.adsabs.harvard.edu/abs/2023ApJ...954L..46L} {954, L46}

\bibitem[\protect\citeauthoryear{{Lewis} \& {Challinor}}{{Lewis} \&
  {Challinor}}{2011}]{Lewis..2011}
{Lewis} A.,  {Challinor} A.,  2011, {CAMB: Code for Anisotropies in the
  Microwave Background}, Astrophysics Source Code Library, record ascl:1102.026

\bibitem[\protect\citeauthoryear{{Lin} et~al.,}{{Lin} et~al.}{2026}]{Lin..2026}
{Lin} X.,  et~al., 2026, \mn@doi [\apj] {10.3847/1538-4357/ae225e}, \href
  {https://ui.adsabs.harvard.edu/abs/2026ApJ...997..207L} {997, 207}

\bibitem[\protect\citeauthoryear{{Liu} \& {Bromm}}{{Liu} \&
  {Bromm}}{2022}]{Liu..2022}
{Liu} B.,  {Bromm} V.,  2022, \mn@doi [\apjl] {10.3847/2041-8213/ac927f}, \href
  {https://ui.adsabs.harvard.edu/abs/2022ApJ...937L..30L} {937, L30}

\bibitem[\protect\citeauthoryear{Livermore, Finkelstein  \& Lotz}{Livermore
  et~al.}{2017}]{Livermore..2017}
Livermore R.~C.,  Finkelstein S.~L.,   Lotz J.~M.,  2017, \mn@doi [\apj]
  {10.3847/1538-4357/835/2/113}, 835, 113

\bibitem[\protect\citeauthoryear{{Mason}, {Trenti}  \& {Treu}}{{Mason}
  et~al.}{2023}]{Mason..2023}
{Mason} C.~A.,  {Trenti} M.,   {Treu} T.,  2023, \mn@doi [\mnras]
  {10.1093/mnras/stad035}, \href
  {https://ui.adsabs.harvard.edu/abs/2023MNRAS.521..497M} {521, 497}

\bibitem[\protect\citeauthoryear{{McLeod} et~al.,}{{McLeod}
  et~al.}{2024}]{Mcleod..2024}
{McLeod} D.~J.,  et~al., 2024, \mn@doi [\mnras] {10.1093/mnras/stad3471}, \href
  {https://ui.adsabs.harvard.edu/abs/2024MNRAS.527.5004M} {527, 5004}

\bibitem[\protect\citeauthoryear{{Mehta} et~al.,}{{Mehta}
  et~al.}{2017}]{Mehta..2017}
{Mehta} V.,  et~al., 2017, \mn@doi [\apj] {10.3847/1538-4357/aa6259}, \href
  {https://ui.adsabs.harvard.edu/abs/2017ApJ...838...29M} {838, 29}

\bibitem[\protect\citeauthoryear{{Messa} et~al.,}{{Messa}
  et~al.}{2025}]{Messa..2025}
{Messa} M.,  et~al., 2025, \mn@doi [arXiv e-prints]
  {10.48550/arXiv.2507.18705}, \href
  {https://ui.adsabs.harvard.edu/abs/2025arXiv250718705M} {p. arXiv:2507.18705}

\bibitem[\protect\citeauthoryear{{Mu{\~n}oz} et~al.,}{{Mu{\~n}oz}
  et~al.}{2026}]{Munoz..2026}
{Mu{\~n}oz} J.~B.,  et~al., 2026, \mn@doi [\mnras] {10.1093/mnras/stag415},
  \href {https://ui.adsabs.harvard.edu/abs/2026MNRAS.tmp..405M} {}

\bibitem[\protect\citeauthoryear{{Naidu} et~al.,}{{Naidu}
  et~al.}{2022}]{Naidu..2022}
{Naidu} R.~P.,  et~al., 2022, \mn@doi [\apjl] {10.3847/2041-8213/ac9b22}, \href
  {https://ui.adsabs.harvard.edu/abs/2022ApJ...940L..14N} {940, L14}

\bibitem[\protect\citeauthoryear{{Oesch}, {Bouwens}, {Illingworth}, {Labb{\'e}}
   \& {Stefanon}}{{Oesch} et~al.}{2018}]{Oesch..2018}
{Oesch} P.~A.,  {Bouwens} R.~J.,  {Illingworth} G.~D.,  {Labb{\'e}} I.,
  {Stefanon} M.,  2018, \mn@doi [\apj] {10.3847/1538-4357/aab03f}, \href
  {https://ui.adsabs.harvard.edu/abs/2018ApJ...855..105O} {855, 105}

\bibitem[\protect\citeauthoryear{{Oke} \& {Gunn}}{{Oke} \&
  {Gunn}}{1983}]{Oke..1983}
{Oke} J.~B.,  {Gunn} J.~E.,  1983, \mn@doi [\apj] {10.1086/160817}, \href
  {https://ui.adsabs.harvard.edu/abs/1983ApJ...266..713O} {266, 713}

\bibitem[\protect\citeauthoryear{{Ono} et~al.,}{{Ono} et~al.}{2018}]{Ono..2018}
{Ono} Y.,  et~al., 2018, \mn@doi [\pasj] {10.1093/pasj/psx103}, \href
  {https://ui.adsabs.harvard.edu/abs/2018PASJ...70S..10O} {70, S10}

\bibitem[\protect\citeauthoryear{Ostriker \& McKee}{Ostriker \&
  McKee}{1988}]{Ostriker..1988}
Ostriker J.~P.,  McKee C.~F.,  1988, \mn@doi [Rev. Mod. Phys.]
  {10.1103/RevModPhys.60.1}, 60, 1

\bibitem[\protect\citeauthoryear{{Parsa}, {Dunlop}, {McLure}  \&
  {Mortlock}}{{Parsa} et~al.}{2016}]{Parsa..2016}
{Parsa} S.,  {Dunlop} J.~S.,  {McLure} R.~J.,   {Mortlock} A.,  2016, \mn@doi
  [\mnras] {10.1093/mnras/stv2857}, \href
  {https://ui.adsabs.harvard.edu/abs/2016MNRAS.456.3194P} {456, 3194}

\bibitem[\protect\citeauthoryear{{P{\'e}rez-Gonz{\'a}lez}
  et~al.,}{{P{\'e}rez-Gonz{\'a}lez} et~al.}{2023}]{Perez..2023}
{P{\'e}rez-Gonz{\'a}lez} P.~G.,  et~al., 2023, \mn@doi [\apjl]
  {10.3847/2041-8213/acd9d0}, \href
  {https://ui.adsabs.harvard.edu/abs/2023ApJ...951L...1P} {951, L1}

\bibitem[\protect\citeauthoryear{Peterson}{Peterson}{1997}]{Peterson..1997}
Peterson B.~M.,  1997, {An Introduction to Active Galactic Nuclei}.
Cambridge University Press

\bibitem[\protect\citeauthoryear{{Planck Collaboration} et~al.,}{{Planck
  Collaboration} et~al.}{2020}]{Planck..2018}
{Planck Collaboration} et~al., 2020, \mn@doi [\aap]
  {10.1051/0004-6361/201833910}, \href
  {https://ui.adsabs.harvard.edu/abs/2020A&A...641A...6P} {641, A6}

\bibitem[\protect\citeauthoryear{{Reddy} \& {Steidel}}{{Reddy} \&
  {Steidel}}{2009}]{Reddy..2009}
{Reddy} N.~A.,  {Steidel} C.~C.,  2009, \mn@doi [\apj]
  {10.1088/0004-637X/692/1/778}, \href
  {https://ui.adsabs.harvard.edu/abs/2009ApJ...692..778R} {692, 778}

\bibitem[\protect\citeauthoryear{{Richards} et~al.,}{{Richards}
  et~al.}{2006}]{Richards..2006}
{Richards} G.~T.,  et~al., 2006, \mn@doi [\apjs] {10.1086/506525}, \href
  {https://ui.adsabs.harvard.edu/abs/2006ApJS..166..470R} {166, 470}

\bibitem[\protect\citeauthoryear{Robertson et~al.,}{Robertson
  et~al.}{2024}]{Robertson..2024}
Robertson B.,  et~al., 2024, \mn@doi [\apj] {10.3847/1538-4357/ad463d}, 970, 31

\bibitem[\protect\citeauthoryear{Robitaille \& Whitney}{Robitaille \&
  Whitney}{2010}]{Robitaille..2010}
Robitaille T.,  Whitney B.,  2010, \mn@doi [ApJ] {10.1088/2041-8205/710/1/L11},
  710

\bibitem[\protect\citeauthoryear{{Rojas-Ruiz} et~al.,}{{Rojas-Ruiz}
  et~al.}{2025}]{Rojas-Ruiz..2025}
{Rojas-Ruiz} S.,  et~al., 2025, \mn@doi [arXiv e-prints]
  {10.48550/arXiv.2507.01014}, \href
  {https://ui.adsabs.harvard.edu/abs/2025arXiv250701014R} {p. arXiv:2507.01014}

\bibitem[\protect\citeauthoryear{{Runnoe}, {Brotherton}  \& {Shang}}{{Runnoe}
  et~al.}{2012}]{Runnoe..2012}
{Runnoe} J.~C.,  {Brotherton} M.~S.,   {Shang} Z.,  2012, \mn@doi [\mnras]
  {10.1111/j.1365-2966.2012.20620.x}, \href
  {https://ui.adsabs.harvard.edu/abs/2012MNRAS.422..478R} {422, 478}

\bibitem[\protect\citeauthoryear{Rybicki \& Lightman}{Rybicki \&
  Lightman}{1979}]{Rybicki..1979}
Rybicki G.~B.,  Lightman A.~P.,  1979, {Radiative Processes in Astrophysics}.
John Wiley \& Sons

\bibitem[\protect\citeauthoryear{{Salpeter}}{{Salpeter}}{1955}]{Salpeter..1955}
{Salpeter} E.~E.,  1955, \mn@doi [\apj] {10.1086/145971}, \href
  {https://ui.adsabs.harvard.edu/abs/1955ApJ...121..161S} {121, 161}

\bibitem[\protect\citeauthoryear{{Samui}}{{Samui}}{2014}]{Samui..2014}
{Samui} S.,  2014, \mn@doi [\na] {10.1016/j.newast.2014.01.010}, \href
  {https://ui.adsabs.harvard.edu/abs/2014NewA...30...89S} {30, 89}

\bibitem[\protect\citeauthoryear{Samui, Srianand  \& Subramanian}{Samui
  et~al.}{2007}]{Samui..2007}
Samui S.,  Srianand R.,   Subramanian K.,  2007, \mn@doi [\mnras]
  {10.1111/j.1365-2966.2007.11603.x}, 377, 285–299

\bibitem[\protect\citeauthoryear{{Samui}, {Subramanian}  \& {Srianand}}{{Samui}
  et~al.}{2008}]{Samui..2008}
{Samui} S.,  {Subramanian} K.,   {Srianand} R.,  2008, \mn@doi [\mnras]
  {10.1111/j.1365-2966.2008.12932.x}, \href
  {https://ui.adsabs.harvard.edu/abs/2008MNRAS.385..783S} {385, 783}

\bibitem[\protect\citeauthoryear{{Samui}, {Srianand}  \& {Subramanian}}{{Samui}
  et~al.}{2018}]{Samui..2018}
{Samui} S.,  {Srianand} R.,   {Subramanian} K.,  2018, \mn@doi [arXiv e-prints]
  {10.48550/arXiv.1805.05945}, \href
  {https://ui.adsabs.harvard.edu/abs/2018arXiv180505945S} {p. arXiv:1805.05945}

\bibitem[\protect\citeauthoryear{{Schmidt}}{{Schmidt}}{1959}]{Schmidt..1959}
{Schmidt} M.,  1959, \mn@doi [\apj] {10.1086/146614}, \href
  {https://ui.adsabs.harvard.edu/abs/1959ApJ...129..243S} {129, 243}

\bibitem[\protect\citeauthoryear{{Shen}, {Vogelsberger}, {Boylan-Kolchin},
  {Tacchella}  \& {Kannan}}{{Shen} et~al.}{2023}]{Shen..2023}
{Shen} X.,  {Vogelsberger} M.,  {Boylan-Kolchin} M.,  {Tacchella} S.,
  {Kannan} R.,  2023, \mn@doi [\mnras] {10.1093/mnras/stad2508}, \href
  {https://ui.adsabs.harvard.edu/abs/2023MNRAS.525.3254S} {525, 3254}

\bibitem[\protect\citeauthoryear{{Sheth} \& {Tormen}}{{Sheth} \&
  {Tormen}}{1999}]{Sheth-Tormen..1999}
{Sheth} R.~K.,  {Tormen} G.,  1999, \mn@doi [\mnras]
  {10.1046/j.1365-8711.1999.02692.x}, \href
  {https://ui.adsabs.harvard.edu/abs/1999MNRAS.308..119S} {308, 119}

\bibitem[\protect\citeauthoryear{{Steinhardt}, {Kokorev}, {Rusakov}, {Garcia}
  \& {Sneppen}}{{Steinhardt} et~al.}{2023}]{Steinhardt..2023}
{Steinhardt} C.~L.,  {Kokorev} V.,  {Rusakov} V.,  {Garcia} E.,   {Sneppen} A.,
   2023, \mn@doi [\apjl] {10.3847/2041-8213/acdef6}, \href
  {https://ui.adsabs.harvard.edu/abs/2023ApJ...951L..40S} {951, L40}

\bibitem[\protect\citeauthoryear{{Sun}, {Faucher-Gigu{\`e}re}, {Hayward}  \&
  {Shen}}{{Sun} et~al.}{2023a}]{sun..2023a}
{Sun} G.,  {Faucher-Gigu{\`e}re} C.-A.,  {Hayward} C.~C.,   {Shen} X.,  2023a,
  \mn@doi [\mnras] {10.1093/mnras/stad2902}, \href
  {https://ui.adsabs.harvard.edu/abs/2023MNRAS.526.2665S} {526, 2665}

\bibitem[\protect\citeauthoryear{{Sun}, {Faucher-Gigu{\`e}re}, {Hayward},
  {Shen}, {Wetzel}  \& {Cochrane}}{{Sun} et~al.}{2023b}]{sun..2023b}
{Sun} G.,  {Faucher-Gigu{\`e}re} C.-A.,  {Hayward} C.~C.,  {Shen} X.,  {Wetzel}
  A.,   {Cochrane} R.~K.,  2023b, \mn@doi [\apjl] {10.3847/2041-8213/acf85a},
  \href {https://ui.adsabs.harvard.edu/abs/2023ApJ...955L..35S} {955, L35}

\bibitem[\protect\citeauthoryear{{Sun}, {Mu{\~n}oz}, {Mirocha}  \&
  {Faucher-Gigu{\`e}re}}{{Sun} et~al.}{2025}]{Sun..2025}
{Sun} G.,  {Mu{\~n}oz} J.~B.,  {Mirocha} J.,   {Faucher-Gigu{\`e}re} C.-A.,
  2025, \mn@doi [\jcap] {10.1088/1475-7516/2025/04/034}, \href
  {https://ui.adsabs.harvard.edu/abs/2025JCAP...04..034S} {2025, 034}

\bibitem[\protect\citeauthoryear{{Tacchella} et~al.,}{{Tacchella}
  et~al.}{2022}]{Tacchella..2022}
{Tacchella} S.,  et~al., 2022, \mn@doi [\apj] {10.3847/1538-4357/ac4cad}, \href
  {https://ui.adsabs.harvard.edu/abs/2022ApJ...927..170T} {927, 170}

\bibitem[\protect\citeauthoryear{{Tremonti} et~al.,}{{Tremonti}
  et~al.}{2004}]{Tremonti..2004}
{Tremonti} C.~A.,  et~al., 2004, \mn@doi [\apj] {10.1086/423264}, \href
  {https://ui.adsabs.harvard.edu/abs/2004ApJ...613..898T} {613, 898}

\bibitem[\protect\citeauthoryear{{Trenti} et~al.,}{{Trenti}
  et~al.}{2011}]{BoRG..2011}
{Trenti} M.,  et~al., 2011, \mn@doi [\apjl] {10.1088/2041-8205/727/2/L39},
  \href {https://ui.adsabs.harvard.edu/abs/2011ApJ...727L..39T} {727, L39}

\bibitem[\protect\citeauthoryear{{Tumlinson}}{{Tumlinson}}{2006}]{Tumilson..2006}
{Tumlinson} J.,  2006, \mn@doi [\apj] {10.1086/500383}, \href
  {https://ui.adsabs.harvard.edu/abs/2006ApJ...641....1T} {641, 1}

\bibitem[\protect\citeauthoryear{{Vallini} et~al.,}{{Vallini}
  et~al.}{2024}]{Vallini..2024}
{Vallini} L.,  et~al., 2024, \mn@doi [\mnras] {10.1093/mnras/stad3150}, \href
  {https://ui.adsabs.harvard.edu/abs/2024MNRAS.527...10V} {527, 10}

\bibitem[\protect\citeauthoryear{{Vasudevan} \& {Fabian}}{{Vasudevan} \&
  {Fabian}}{2007}]{Vasudevan..2007}
{Vasudevan} R.~V.,  {Fabian} A.~C.,  2007, \mn@doi [\mnras]
  {10.1111/j.1365-2966.2007.12328.x}, \href
  {https://ui.adsabs.harvard.edu/abs/2007MNRAS.381.1235V} {381, 1235}

\bibitem[\protect\citeauthoryear{{Veilleux}, {Cecil}  \&
  {Bland-Hawthorn}}{{Veilleux} et~al.}{2005}]{Veilleux..2005}
{Veilleux} S.,  {Cecil} G.,   {Bland-Hawthorn} J.,  2005, \mn@doi [\araa]
  {10.1146/annurev.astro.43.072103.150610}, \href
  {https://ui.adsabs.harvard.edu/abs/2005ARA&A..43..769V} {43, 769}

\bibitem[\protect\citeauthoryear{{Weaver}, {McCray}, {Castor}, {Shapiro}  \&
  {Moore}}{{Weaver} et~al.}{1977}]{Weaver..1977}
{Weaver} R.,  {McCray} R.,  {Castor} J.,  {Shapiro} P.,   {Moore} R.,  1977,
  \mn@doi [\apj] {10.1086/155692}, \href
  {https://ui.adsabs.harvard.edu/abs/1977ApJ...218..377W} {218, 377}

\bibitem[\protect\citeauthoryear{{Weibel} et~al.,}{{Weibel}
  et~al.}{2025}]{Weibel..2025}
{Weibel} A.,  et~al., 2025, \mn@doi [\apj] {10.3847/1538-4357/adab7a}, \href
  {https://ui.adsabs.harvard.edu/abs/2025ApJ...983...11W} {983, 11}

\bibitem[\protect\citeauthoryear{{Wyithe} \& {Padmanabhan}}{{Wyithe} \&
  {Padmanabhan}}{2006}]{Stuart..2006}
{Wyithe} J. S.~B.,  {Padmanabhan} T.,  2006, \mn@doi [\mnras]
  {10.1111/j.1365-2966.2005.09858.x}, \href
  {https://ui.adsabs.harvard.edu/abs/2006MNRAS.366.1029W} {366, 1029}

\bibitem[\protect\citeauthoryear{{Yung}, {Somerville}, {Finkelstein}, {Wilkins}
   \& {Gardner}}{{Yung} et~al.}{2024}]{Yung..2024}
{Yung} L.~Y.~A.,  {Somerville} R.~S.,  {Finkelstein} S.~L.,  {Wilkins} S.~M.,
  {Gardner} J.~P.,  2024, \mn@doi [\mnras] {10.1093/mnras/stad3484}, \href
  {https://ui.adsabs.harvard.edu/abs/2024MNRAS.527.5929Y} {527, 5929}

\bibitem[\protect\citeauthoryear{{van den Bosch}, {Aquino}, {Yang}, {Mo},
  {Pasquali}, {McIntosh}, {Weinmann}  \& {Kang}}{{van den Bosch}
  et~al.}{2008}]{vdb..2008}
{van den Bosch} F.~C.,  {Aquino} D.,  {Yang} X.,  {Mo} H.~J.,  {Pasquali} A.,
  {McIntosh} D.~H.,  {Weinmann} S.~M.,   {Kang} X.,  2008, \mn@doi [\mnras]
  {10.1111/j.1365-2966.2008.13230.x}, \href
  {https://ui.adsabs.harvard.edu/abs/2008MNRAS.387...79V} {387, 79}

\bibitem[\protect\citeauthoryear{{van der Burg}, {Hildebrandt}  \&
  {Erben}}{{van der Burg} et~al.}{2010}]{vdb..2010}
{van der Burg} R.~F.~J.,  {Hildebrandt} H.,   {Erben} T.,  2010, \mn@doi [\aap]
  {10.1051/0004-6361/200913812}, \href
  {https://ui.adsabs.harvard.edu/abs/2010A&A...523A..74V} {523, A74}

\makeatother
\end{thebibliography}

%\begin{thebibliography}{99}
%\end{thebibliography}

%%%%%%%%%%%%%%%%%%%%%%%%%%%%%%%%%%%%%%%%%%%%%%%%%%

%%%%%%%%%%%%%%%%% APPENDICES %%%%%%%%%%%%%%%%%%%%%

\appendix
\section{Variation of star formation with model parameters} \label{sfh_kappa_relation}

\begin{figure*}
\centerline{
\includegraphics[width=\textwidth]{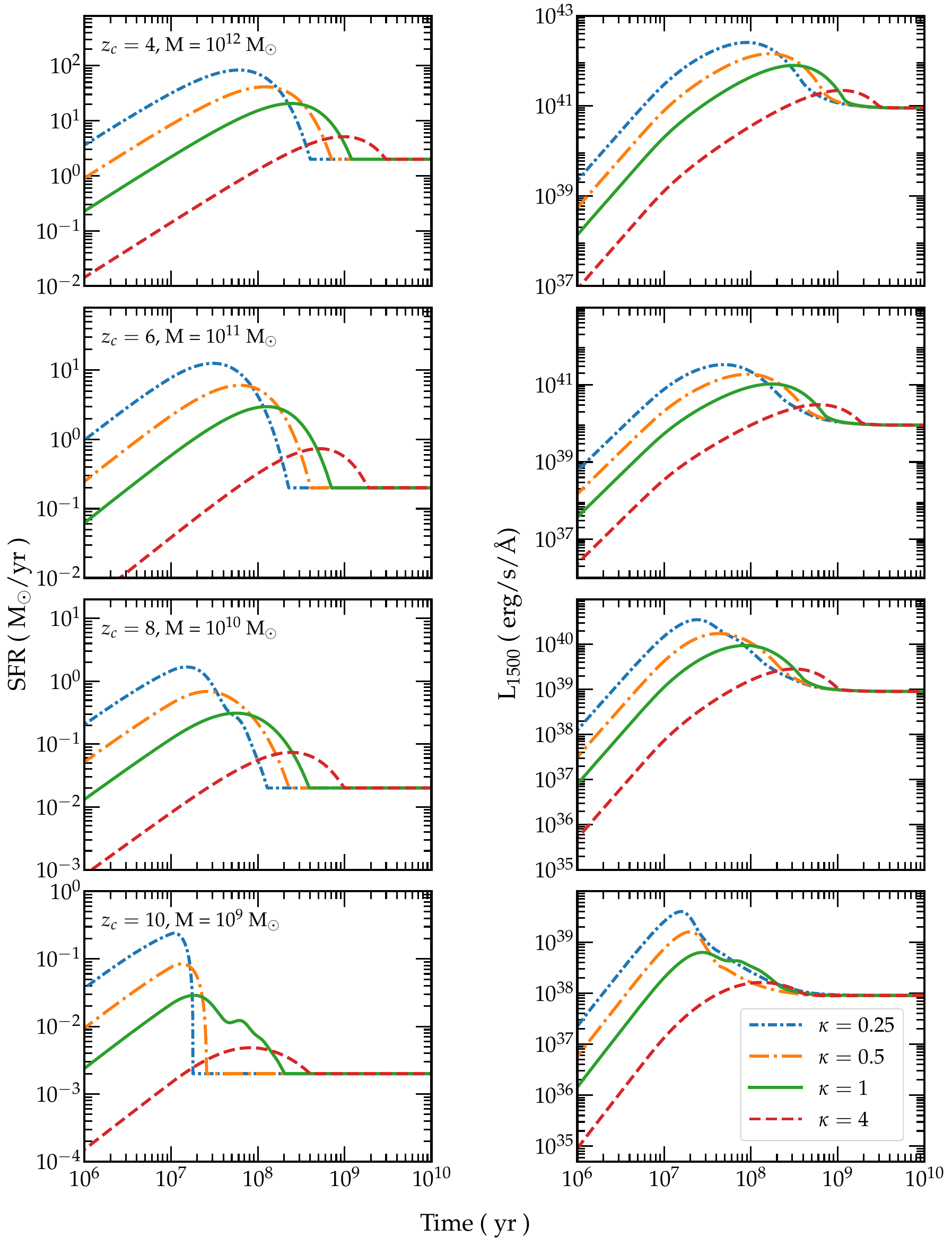}%

}
\caption[]{The evolution of the star formation rate (SFR) and the corresponding UV luminosity ($L_{1500}$) as functions of galaxy age for halo masses of $10^9$, $10^{10}$, $10^{11}$, and $10^{12}$~M$_\odot$. The left panel shows the SFR as a function of age, while the right panel presents the corresponding UV luminosity. The dashed red, solid green, long dash-dotted orange, and short dash-dotted blue curves represent models with $\kappa = 4$, 1, 0.5, and 0.25, respectively.}

\label{kappa_comparison}
\end{figure*}

As mentioned in the main text, we have a few star formation related model parameters, like $f_*$, $\kappa$, IMF etc. In addition, it was shown that the evolution of $\kappa$ plays the most important role in explaining the observed luminosity functions over the entire observable range of $z=2-14$. Thus in this appendix we show how the star formation rate and the resulting UV luminosity ($L_{1500}$) of individual galaxies change with different values of $\kappa$ that are important in our model. In Figure~\ref{kappa_comparison}, we present the star formation rate (SFR; left panels) and the UV luminosity at 1500~\AA~ (right panels) for galaxies with halo masses of $10^{12}$, $10^{11}$, $10^{10}$, $10^{9}$, and $10^{8}$~M$_\odot$ (from top to bottom) collapsed at redshifts of $z_c = 4, 6, 8, 10,$ and $12$, respectively. We select $z_c$ such that each of these haloes is likely to collapse from a $3\sigma$ density fluctuation at that redshift.

As expected, the effect of $\kappa$ is to regulate how rapidly the available cold gas is converted into stars. Smaller values of $\kappa$ correspond to shorter star formation time scales, so the gas is consumed more rapidly, producing an earlier and more pronounced peak in both SFR and $L_{1500}$. In contrast, larger $\kappa$ values delay the onset of efficient star formation and spread the activity over a longer period, resulting in a lower peak amplitude but a more extended evolutionary phase.

We note that this behavior is a direct consequence of the star formation timescale in our model. For example, in a halo of mass $10^{12}\,\rm M_\odot$, the model with $\kappa = 0.25$ reaches a peak SFR of $\sim 100$~M$_\odot$~yr$^{-1}$ and $L_{1500} \sim 3 \times 10^{42}$~erg~s$^{-1}$\AA$^{-1}$, whereas for $\kappa = 4$, the corresponding values are $\sim 50\,\rm M_\odot\,{\rm yr}^{-1}$ and $\sim 3 \times 10^{40}$~erg~s$^{-1}$\AA$^{-1}$.
This is because a lower $\kappa$ leads to a shorter star formation timescale, allowing gas to be converted into stars more rapidly, thereby enhancing the instantaneous UV output. In contrast, higher $\kappa$ values distribute star formation over longer timescales, resulting in lower instantaneous SFR and UV luminosity. As a consequence, models with higher $\kappa$ require a larger star formation efficiency to reproduce the same observed UV luminosity function, while models with lower $\kappa$ naturally achieve higher luminosities and therefore prefer lower efficiencies.

This behavior is seen in all halo masses considered here, although the exact peak height and peak time depend on the halo mass as the mass-dependent supernova feedback regulates the peak star formation rate. Thus for a smaller $\kappa$, a galaxy would be visible to us early in time and would be observed as a brighter galaxy; however, it would remain observable (given a magnitude-limited observation) for a shorter period of time. On the other hand, a large $\kappa$ would lead to a less bright galaxy but that galaxy remains observable for a longer period. This, in turn, changes the slope of the luminosity functions for different values of $\kappa$ as can be seen from Figure~\ref{fig:LF:2-7}. Further, more massive halos reach higher SFRs and UV luminosities due to the availability of more gas for star formation, whereas lower-mass galaxies show weaker and shorter-lived activity owing to stronger supernova feedback, as can be seen in the bottom panel of Figure~\ref{kappa_comparison}.

\begin{figure}
\centerline{
\includegraphics[width=\columnwidth]{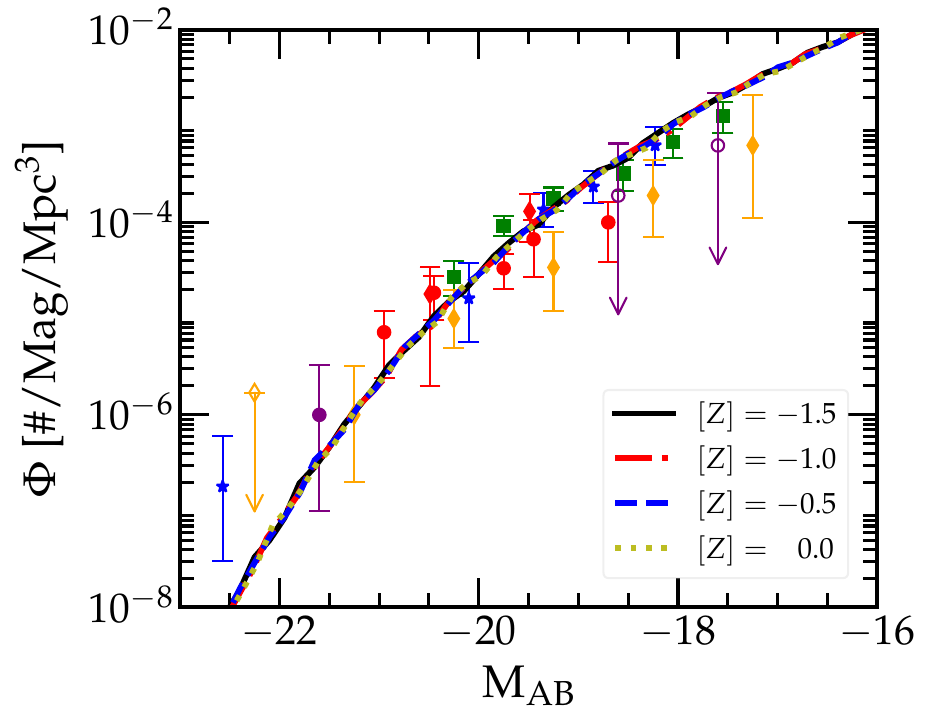}%

}
\caption[]{The figure illustrates the effect of varying metallicity on the nature of the UV LF at $z \sim 10$ (fitted with the observational data). The black solid line represents our fiducial model with $\left[ Z \right] = -1.5$, while the dash-dotted red, dashed blue, and dotted olive lines correspond to $\left[ Z \right] = -1.0$, $-0.5$, and $0.0$ (solar metallicity), respectively.}

\label{Metalicity Comparison}
\end{figure}

\section{Effect of Metallicity on UV LF} \label{Effect of metalicity}

In our fiducial model, we have assumed a metallicity of $\left[ Z \right] = -1.5$ to construct the UV LF. Here, we investigate the impact of varying metallicity on the UV LF. 
In Figure~\ref{Metalicity Comparison}, we present the best fit UV LF at $z = 10$ for four different metallicities: $\left[ Z \right] = -1.5, -1.0, -0.5,$ and $0.0$, where $0.0$ corresponds to solar metallicity. It is clear from the figure that overall nature of the UV LF remains unchanged across this range of metallicities. All models provide a similar fit to the observed luminosity function as confirmed from the value of the reduced chi-square. In particular, the values of reduced chi-square are 1.507, 1.618, 1.619 and 1.612 for $\left[ Z \right] = -1.5, -1.0, -0.5,$ and $0.0$, respectively. This is achieved with only a slight increase in the star formation efficiency parameter, $f_\star$, with increasing metallicity. Specifically, $f_\star$ increases from $\sim 0.07$ at $\left[ Z \right] = -1.5$ to $\sim 0.08$ at $\left[ Z \right] = -0.5$, and further to $\sim 0.09$ at solar metallicity. Thus, the effect of metallicity is relatively modest on the resulting UV luminosity functions and adopting a fixed metallicity is a reasonable approximation for modeling the high-redshift UV LF. A similar trend has been seen in other redshifts as well, which are not shown here. We further note that a model with metallicity variation across the halo mass as observed in galaxies (i.e. mass-metallicity relation) would be ideal. However, the observed metallicity variation across the halo mass range is very small i.e. metallicity changes about 0.5 dex for stellar mass range of $10^8-10^{11}$~M$_\odot$ \citep{Tremonti..2004, Andrews..2013,Curti..2020}. Therefore, we expect a negligible effect on the UV luminosity with such a detailed model which is beyond the scope of this paper.

%%%%%%%%%%%%%%%%%%%%%%%%%%%%%%%%%%%%%%%%%%%%%%%%%%%%%%%%%%%%%%%%%%%%%%%%%%%%%%%%%%%%%%

% Don't change these lines
\bsp	% typesetting comment
\label{lastpage}
\end{document}